\documentclass[iop]{emulateapj}
\usepackage{graphicx}
\usepackage{enumerate}
\usepackage{amssymb, amsmath}
\usepackage{natbib}
\usepackage{color}
\usepackage{ulem}
\usepackage{dblfnote}
\usepackage{appendix}
\usepackage{hyperref}
\usepackage[stable]{footmisc}
\usepackage{float}
\usepackage{comment}
\usepackage{footnote}

\begin{document}

\newcommand{\ipac}{1}
\newcommand{\nexsci}{2}
\newcommand{\naoj}{3}
\newcommand{\kogakuin}{4}
\newcommand{\caltech}{5}
\newcommand{\jpl}{6}
\newcommand{\monarsh}{7}
\newcommand{\abc}{8}
\newcommand{\subaru}{9}
\newcommand{\sokendai}{10}
\newcommand{\liege}{11}
\newcommand{\hokkaido}{12}
\newcommand{\carnegie}{13}
\newcommand{\aix}{14}
\newcommand{\ames}{15}
\newcommand{\eureka}{16}
\newcommand{\victoria}{17}
\newcommand{\amherst}{18}
\newcommand{\zurich}{19}
\newcommand{\lesia}{20}
\newcommand{\tokyo}{21}
\newcommand{\sokendaihayama}{22}
\newcommand{\leuven}{23}
\newcommand{\northridge}{24}
\newcommand{\eso}{25}
\newcommand{\caltechgps}{26}
\newcommand{\uhmanoa}{27}
\newcommand{\lasvegas}{28}

\newcommand{\rdnote}[1]{{\bf \color{red}[R.D.: #1]}}

\author{Taichi Uyama$^{\ipac,\nexsci,\naoj}$}
\author{Takayuki Muto$^{\kogakuin}$}
\author{Dimitri Mawet$^{\caltech,\jpl}$}
\author{Valentin Christiaens$^{\monarsh}$}
\author{Jun Hashimoto$^{\abc,\subaru,\sokendai}$}
\author{Tomoyuki Kudo$^{\subaru}$}
\author{Masayuki Kuzuhara$^{\abc}$}
\author{Garreth Ruane$^{\jpl}$}
\author{Charles Beichman$^{\ipac,\nexsci}$}
\author{Olivier Absil$^{\liege}$}
\author{Eiji Akiyama$^{\hokkaido}$}
\author{Jaehan Bae$^{\carnegie}$}
\author{Michael Bottom$^{\jpl}$}
\author{Elodie Choquet$^{\aix}$}
\author{Thayne Currie$^{\ames,\subaru,\eureka}$}
\author{Ruobing Dong$^{\victoria}$}
\author{Katherine B. Follette$^{\amherst}$}
\author{Misato Fukagawa$^{\naoj}$}
\author{Greta Guidi$^{\zurich}$}
\author{Elsa Huby$^{\lesia}$}
\author{Jungmi Kwon$^{\tokyo}$}
\author{Satoshi Mayama$^{\sokendaihayama}$}
\author{Tiffany Meshkat$^{\ipac}$}
\author{Maddalena Reggiani$^{\leuven}$}
\author{Luca Ricci$^{\northridge}$}
\author{Eugene Serabyn$^{\jpl}$}
\author{Motohide Tamura$^{\tokyo,\abc,\naoj}$}
\author{Leonardo Testi$^{\eso}$}
\author{Nicole Wallack$^{\caltechgps}$}
\author{Jonathan Williams$^{\uhmanoa}$}
\author{Zhaohuan Zhu$^{\lasvegas}$}

\footnotetext[\ipac]{Infrared Processing and Analysis Center, California Institute of Technology, 1200 E. California Blvd., Pasadena, CA 91125, USA}
\footnotetext[\nexsci]{NASA Exoplanet Science Institute, Pasadena, CA 91125, USA}
\footnotetext[\naoj]{National Astronomical Observatory of Japan, 2-21-1 Osawa, Mitaka, Tokyo 181-8588, Japan}
\footnotetext[\kogakuin]{Division of Liberal Arts, Kogakuin University, 1-24-2, Nishi-Shinjuku, Shinjuku-ku, Tokyo, 163-8677, Japan}
\footnotetext[\caltech]{Department of Astronomy, California Institute of Technology, 1200 E. California Blvd., Pasadena, CA 91125, USA}
\footnotetext[\jpl]{Jet Propulsion Laboratory, California Institute of Technology, 4800 Oak Grove Dr., Pasadena, CA, 91109, USA}
\footnotetext[\monarsh]{Monash Centre for Astrophysics (MoCA) and School of Physics and Astronomy, Monash University, Clayton Vic 3800, Australia}
\footnotetext[\abc]{Astrobiology Center, National Institutes of Natural Sciences, 2-21-1 Osawa, Mitaka, Tokyo 181-8588, Japan}
\footnotetext[\subaru]{Subaru Telescope, National Astronomical Observatory of Japan, Mitaka, Tokyo, 181-8588, Japan}
\footnotetext[\sokendai]{Department of Astronomy, School of Science, Graduate University for Advanced Studies (SOKENDAI), Mitaka, Tokyo 181-8588, Japan}
\footnotetext[\liege]{Space Sciences, Technologies, and Astrophysics Research (STAR) Institute, Universit$\acute{e}$ de Li$\grave{e}$ge, Li$\grave{e}$ge, Belgium}
\footnotetext[\hokkaido]{Institute for the Advancement of Higher Education, Hokkaido University, Kita17, Nishi8, 
Kita-ku, Sapporo, 060-0817, Japan}
\footnotetext[\carnegie]{Department of Terrestrial Magnetism, Carnegie Institution for Science, 5241 Broad Branch Road NW, Washington, DC 20015, USA}
\footnotetext[\aix]{Aix Marseille Univ, CNRS, CNES, LAM, Marseille, France}
\footnotetext[\ames]{NASA-Ames Research Center, Moffett Field, CA, USA}
\footnotetext[\eureka]{Eureka Scientific, 2452 Delmer Street Suite 100, Oakland, CA, USA}
\footnotetext[\victoria]{Department of Physics \& Astronomy, University of Victoria, Victoria, BC, V8P 1A1, Canada}
\footnotetext[\amherst]{Physics and Astronomy Department, Amherst College, 21 Merrill Science Drive, Amherst, MA 01002, USA}
\footnotetext[\zurich]{ETH Zurich, Institute for Particle Physics and Astrophysics, Wolfgang-Pauli-Str. 27, 8093 Zurich, Switzerland}
\footnotetext[\lesia]{LESIA, Observatoire de Paris, Universit́$\acute{e}$ PSL, CNRS, Sorbonne Universit$́\acute{e}$, Univ. Paris Diderot, Sorbonne Paris Cit$\acute{e}$ ,5 place Jules Janssen, 92195 Meudon, France}
\footnotetext[\tokyo]{Department of Astronomy, The University of Tokyo, 7-3-1, Hongo, Bunkyo-ku, Tokyo 113-0033, Japan}
\footnotetext[\sokendaihayama]{The Graduate University for Advanced Studies, SOKENDAI, Shonan
International Village, Hayama-cho, Miura-gun, Kanagawa 240-0193, Japan}
\footnotetext[\leuven]{Institute of Astrophysics, KU Leuven, Celestijnlaan 200D, 3001 Leuven, Belgium}
\footnotetext[\northridge]{Department of Physics and Astronomy, California State University Northridge, 18111 Nordhoff Street, Northridge, CA 91330, USA}
\footnotetext[\eso]{ESO/European Southern Observatory, Karl-Schwarzschild-Strasse 2, 85748 Garching bei München, Germany}
\footnotetext[\caltechgps]{Division of Geological and Planetary Sciences, California Institute of Technology, 1200 E. California Blvd., Pasadena, CA 91125, USA}
\footnotetext[\uhmanoa]{Institute for Astronomy, University of Hawaii, Honolulu, HI 96822, USA}
\footnotetext[\lasvegas]{Department of Physics and Astronomy, University of Nevada, Las Vegas, 4505 S. Maryland Pkwy, Las Vegas, NV, 89154-4002}

\title{Near-Infrared Imaging of a Spiral in the CQ Tau Disk}

\begin{abstract}
We present $L^\prime$-band Keck/NIRC2 imaging and $H$-band Subaru/AO188+HiCIAO polarimetric observations of CQ Tau disk with a new spiral arm. Apart from the spiral feature our observations could not detect any companion candidates.
We traced the spiral feature from the $r^2$-scaled HiCIAO polarimetric intensity image and the fitted result is used for forward modeling to reproduce the ADI-reduced NIRC2 image. 
We estimated the original surface brightness after throughput correction in $L^\prime$-band to be $\sim126$ mJy/arcsec$^2$ at most.
We suggest that the grain temperature of the spiral may be heated up to $\sim$200 K in order to explain both of the $H$- and $L^{\prime}$-bands results. The $H$-band emission at the location of the spiral originates from the scattering from the disk surface while both scattering and thermal emission may contribute to the $L^{\prime}$-band emission.
If the central star is only the light source of scattered light, the spiral emission at $L^\prime$-band should be thermal emission.  If an inner disk also acts as the light source, the scattered light and the thermal emission may equally contribute to the $L^\prime$-band spiral structure.

\end{abstract}

\section{Introduction} \label{sec: Introduction} 
Protoplanetary disks are good laboratories for understanding the relationship between planet formation and disk evolution mechanisms.
Previous photometric/spectroscopic studies of young stellar objects (YSOs) with infrared (IR) excesses predicted gaps in their disks \citep[transitional disk;][]{Strom1989}.
As instruments have developed, high-spatial resolution observations with near-IR polarimetric imaging or (sub-)mm interferometry revealed more asymmetric disks with gaps \citep[e.g.,][]{Hashimoto2012}, rings \citep[e.g.,][]{ALMA2015,DSHARP-Andrews2018}, spirals \citep[e.g.,][]{Muto2012,Benisty2015,Perez2016,Uyama2018,DSHARP-Huang2018}, dust traps \citep[e.g.,][]{vanderMarel2013}, asymmetric blob in the disk midplane \citep[e.g.,][]{Tsukagoshi2019}, and velocity kink in gas kinematics \citep[e.g.,][]{Pinte2018}.
In particular, spiral arms are one of the most intriguing signposts of planet formation in the disk because a protoplanet behaves as a perturber of the disk, which can lead to spiral formation \citep[][]{Zhu2015,Dong2018}, but yet no confirmed connection between an observed spiral arm and a planetary mass companion has been made observationally \citep[but see][]{Wagner2019}. Gravitational instability in the disk can produce spirals \citep[][]{Dong2015GI}.

Radio continuum observations measure thermal emission of dust grains in the disk midplane and those at different excitations of gas such as CO(2-1) and CO(3-2) can probe the distribution of different layers of molecular gas species. 
Performing interferometric observation in radio wavelength enables to achieve sufficient spatial resolution to resolve detailed asymmetric structure.
High-contrast broad-band imaging with a variety of differential imaging methods can sometimes detect intriguing disk features.
Polarization differential imaging \citep[PDI;][]{Kuhn2001} provide polarimetric intensity (PI), which traces scattered starlight from the disk surface.
Those explorations with angular differential imaging \citep[ADI;][]{Marois2008} for young planets, in parallel to disk studies, have not successfully detected the most convincing protoplanets within such disks until PDS 70b was reported recently \citep[][]{Keppler2018}.
The results of that paper support the theory that planets really form in protoplanetary disks.
Interestingly, several $L^\prime$-band observations successfully detected asymmetric disk features with ADI \citep[e.g. HD 142527, HD 100546, and MWC 758;][]{Rameau2012, 2015ApJ...814L..27C, Reggiani2018, Wagner2019}.
As \citet{Lyra2016} performed a 3D simulation and predicted that a high-mass planet can induce shocks and heats the spiral to a few hundred Kelvin (see Figure 4 in the paper), the $L^\prime$-band observation has capability to detect thermal emission from the disk.
In the near future one can expect to discover more planets undergoing formation and further searches for asymmetric disk features as well as for protoplanets will help to understand the links between planet formation and disk evolution.

CQ Tau (RA = 05:35:58.47, Dec = +24:44:54.1) is a YSO in the Taurus star forming region \citep[F2-type, $1.67 M_\odot$, $\sim$10 Myr, 162 pc;][]{Natta2001,Gaia2018DR2,Ubeira2019}.
${\rm C_I}$ observations by APEX and comparison with chemical models suggested that CQ Tau likely has a transitional disk \citep[][]{Chapillon2010}. 
An ALMA observation reported a large gap in the 1.3-mm continuum, $^{13}$CO, and $^{18}$CO \citep[][]{Ubeira2019}.
The gap sizes in the dust and gas are estimated at 56 au and 20 au in radius, respectively.
\citet{Ubeira2019} also performed a 3D numerical simulations and suggested an unseen protoplanet in the disk.
To further search for protoplanets as well as asymmetric features in the CQ Tau disk, we used two high-contrast imaging observations with Keck/NIRC2 and Subaru/AO188+HiCIAO.
Although we did not detect any companion candidates, we detected a spiral feature in the disk.
In this study, we analyze the detected spiral feature and investigate the possible links to ongoing planet formation.

\section{Observations and Results} \label{sec: Observations and Results}
We used two infrared data sets taken from Keck/NIRC2 and Subaru/AO188+HiCIAO.
We also used an ALMA archival image, observed in Cycle 5 (ID: 2017.1.01404.S, PI: L. Testi), which achieved a noise level of $\sim$23 $\mu$Jy/beam and a beam size of 69 mas and 51 mas for major-axis and minor-axis, respectively, for comparison with the infrared data. Details of this data set as well as other ALMA data of CQ Tau are described in \citet{Ubeira2019}.
Table \ref{observing logs} summarizes observing logs for both observations. Sections \ref{sec: Keck/NIRC2} and \ref{sec: Subaru/HiCIAO} describe each observation and its result. Section \ref{sec: Commparison of the Two Data Sets} compares both results.

\begin{table*}
    \centering
    \caption{Observing logs}
    \begin{tabular}{cccccc} \hline\hline
    Instrument & Date ({\it UT}) & Observing Mode & Band & Total Exposure Time [sec] \\ \hline
        Keck/NIRC2 & 2018 December 24 & ADI & $L^\prime$ & 1800 \\
        Subaru/HiCIAO & 2015 December 31 & PDI\footnotemark[1] & $H$ & 540 \\
    \end{tabular}
    \label{observing logs}
    \footnotetext[1]{ADI was combined with PDI but we focus on only PDI reduction in this study (see Section \ref{sec: Subaru/HiCIAO}).}
\end{table*}

\subsection{Keck/NIRC2} \label{sec: Keck/NIRC2}
CQ Tau was observed on UT 2018 Dec 24 (PI: D. Mawet) using the Keck/NIRC2 vortex coronagraph \citep[][]{Mawet2017,Serabyn2017,Xuan2018} combined with ADI. The observation achieved angular rotation of $\sim$111$^\circ$.
No standard stars were taken in the same epoch and we did not conduct PSF subtraction by reference differential imaging \citep[RDI;][]{Ruane2019} in this study.
We measured the off-axis PSF and determined that the full width at half maximum (FWHM) was 9.2 pix ($\sim$0\farcs0915 with a pixel scale of 9.972 mas/pix). 
After a first reduction including flat fielding, bad-pixel correction, sky-subtraction, and image registration, the data set was processed via the vortex image processing \citep[VIP;][]{Gonzalez2017}\footnote{\url{https://github.com/vortex-exoplanet}} package that applies principal component analysis (PCA) for the ADI reduction \citep[][]{Amara2012,Soummer2012}.

\begin{figure*}
\begin{tabular}{cc}
 \begin{minipage}{0.5\hsize}
 \centering
 \includegraphics[scale=0.65]{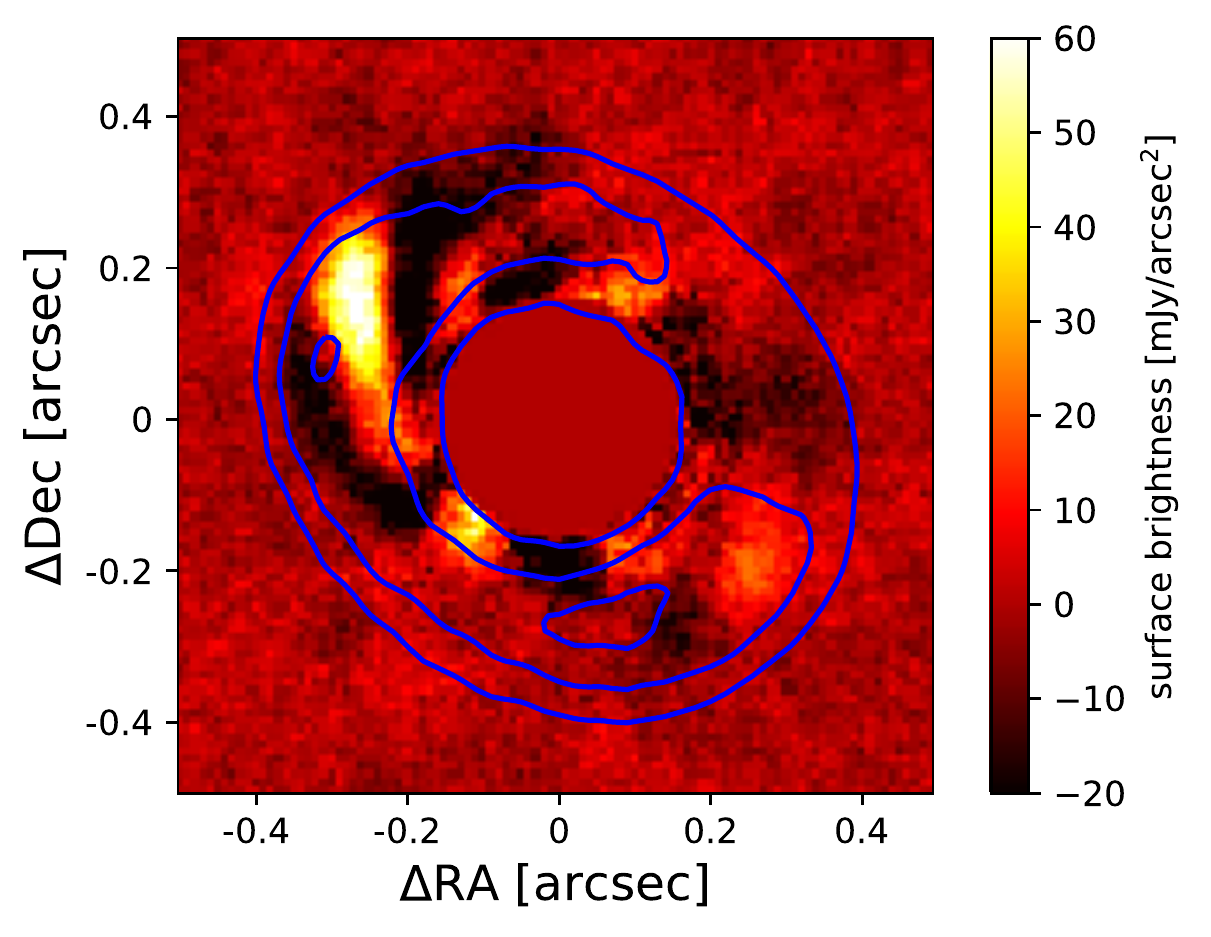}
 \end{minipage}
  \begin{minipage}{0.5\hsize}
 \centering
 \includegraphics[scale=0.65]{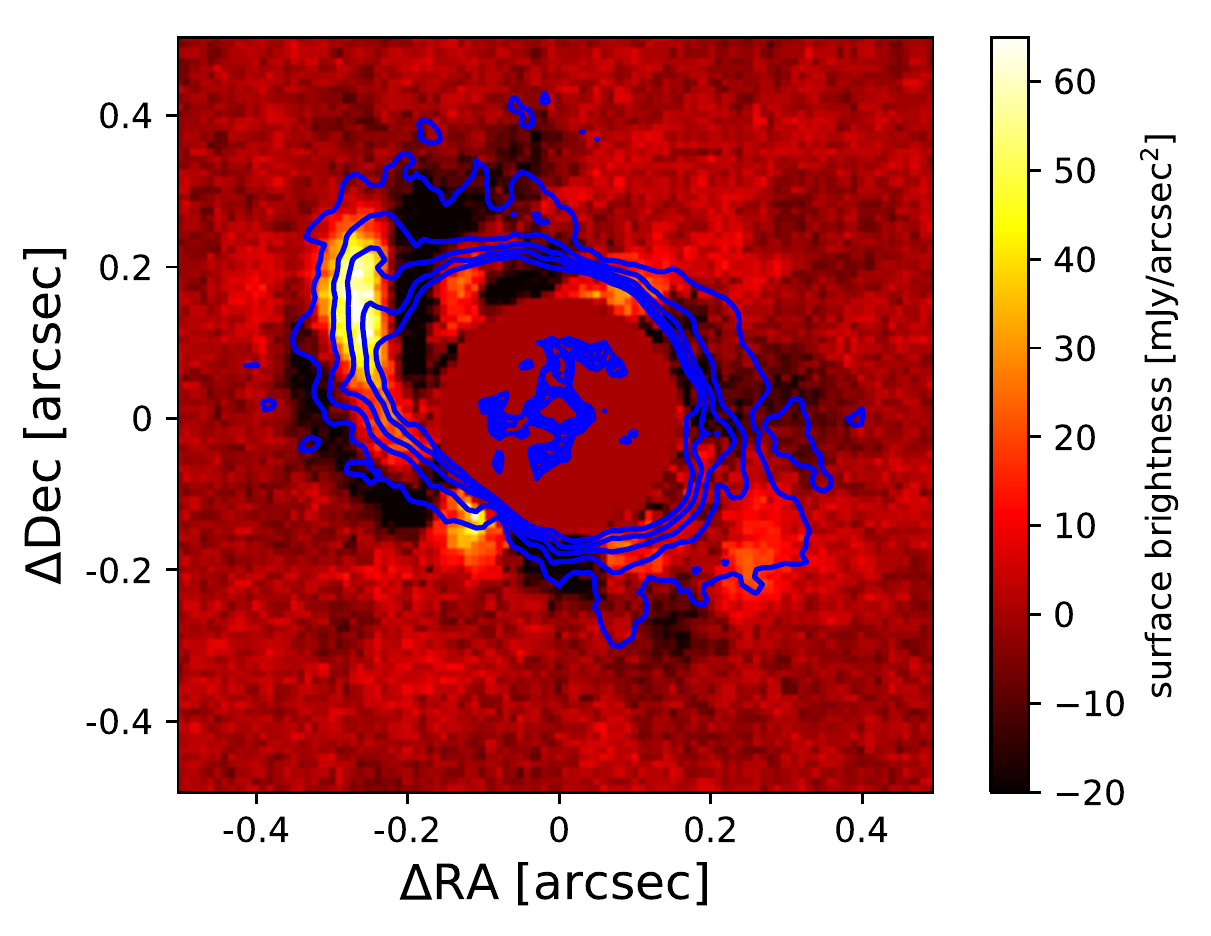}
 \end{minipage}
 \end{tabular}
 \caption{(left) Keck/NIRC2 $L^\prime$-band image of CQ Tau overlaid with contours of the ALMA dust continuum image at 1.3 mm at 30, 50, and 70$\sigma$ (1$\sigma=$23 $\mu$Jy/beam), respectively (blue). The central star is masked by the algorithm. North is up and East is left. The color scale shows surface brightness in mJy/arcsec$^2$ unit. (right) Same as the left figure except for the contours. The contours correspond to Subaru/HiCIAO $H$-band PI data at 10, 20, 30, 40, and 50 mJy/arcsec$^2$, respectively (blue). }
 \label{fig: NIRC2}
\end{figure*}

\begin{figure}
 \centering
 \includegraphics[width=0.48\textwidth]{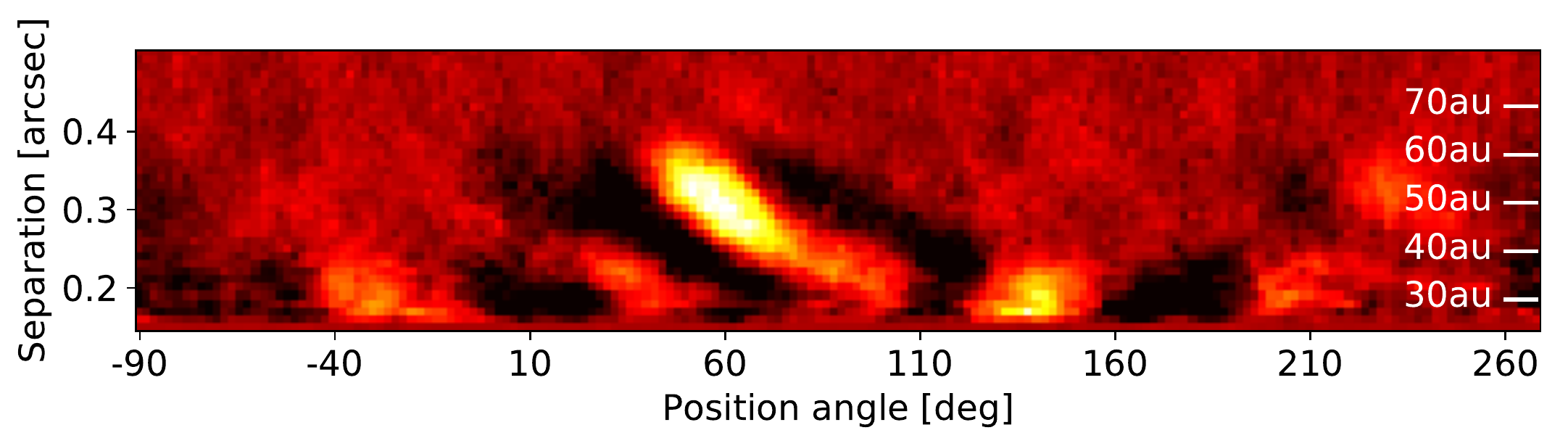}
 \caption{Polar-projected image of the NIRC2 image in Figure \ref{fig: NIRC2}. We arrange the image starting at a position angle of -90$^\circ$ to show the spiral feature clearly.}
 \label{r-theta NIRC2}
\end{figure}

Figure \ref{fig: NIRC2} shows the Keck/NIRC2 ADI-reduced image of CQ Tau overlaid with the ALMA continuum (left) and Subaru/HiCIAO PI (right; see Section \ref{sec: Subaru/HiCIAO} for the data). 
VIP produces a set of different principal components (PCs), results of which are shown in Appendix \ref{sec: Supplementary Keck/NIRC2 Images}. 
We adopted principal component (PC)=8 among these PCs for presenting our result because this image shows an extended object at separations between $\sim$$0\farcs2$ and $0\farcs4$, and position angles (PAs) between $\sim$$45^\circ$ and $110^\circ$ with signal-to-noise ratio (SNR) $\sim$7--8.
The feature appears robust because it survives for a wide range of PC values (see Figure \ref{fig: NIRC2 PCs}).
We marginally found some other sources (see Figure \ref{fig: NIRC2 PCs}) in a set of ADI-reduced images, whose SNRs fall less than 5 at a certain PC and do not discuss other companion candidates.
We converted ADU into the surface brightness using a previous $L^\prime$-band photometry \citep[2.4 Jy for CQ Tau;][]{McDonald2017} and the brightest region in this feature has 68$\pm$8.5 mJy/arcsec$^2$.
The VIP package enables to set different fields of view (FoV) and inner working angle (IWA). 
We adopted IWA=16 pix so that the asymmetric feature is reproduced with a higher SNR.
We reran VIP by setting a smaller IWA to check whether other companion candidates appear at separations smaller than 16 pix and confirmed that there showed only residuals of speckles that vary among different PCs.
We first attempted to fit this extended object with a point-source Gaussian, which provided a poor match and thus we concluded that it corresponds to an asymmetric structure in the CQ Tau disk.
Figure \ref{r-theta NIRC2} shows a polar-projected image suggesting that this feature likely corresponds to a spiral.
CQ Tau is one of only a few systems that have a spiral detected in $L^\prime$-band (see Section \ref{sec: Introduction} for the L'-band disks).

We then compared our results with the ALMA archival data.
The spiral overlaps with the ring of dust continuum, but the ADI-reduced signal experiences self-subtraction by the reduction algorithm as negative regions shown at both sides of the spiral. 
Centrosymmetric features in the CQ Tau disk are also removed by self-subtraction and thus cannot be seen in the ADI-reduced image \citep[][]{Milli2012}.

\begin{figure}
    \centering
    \includegraphics[width=0.48\textwidth]{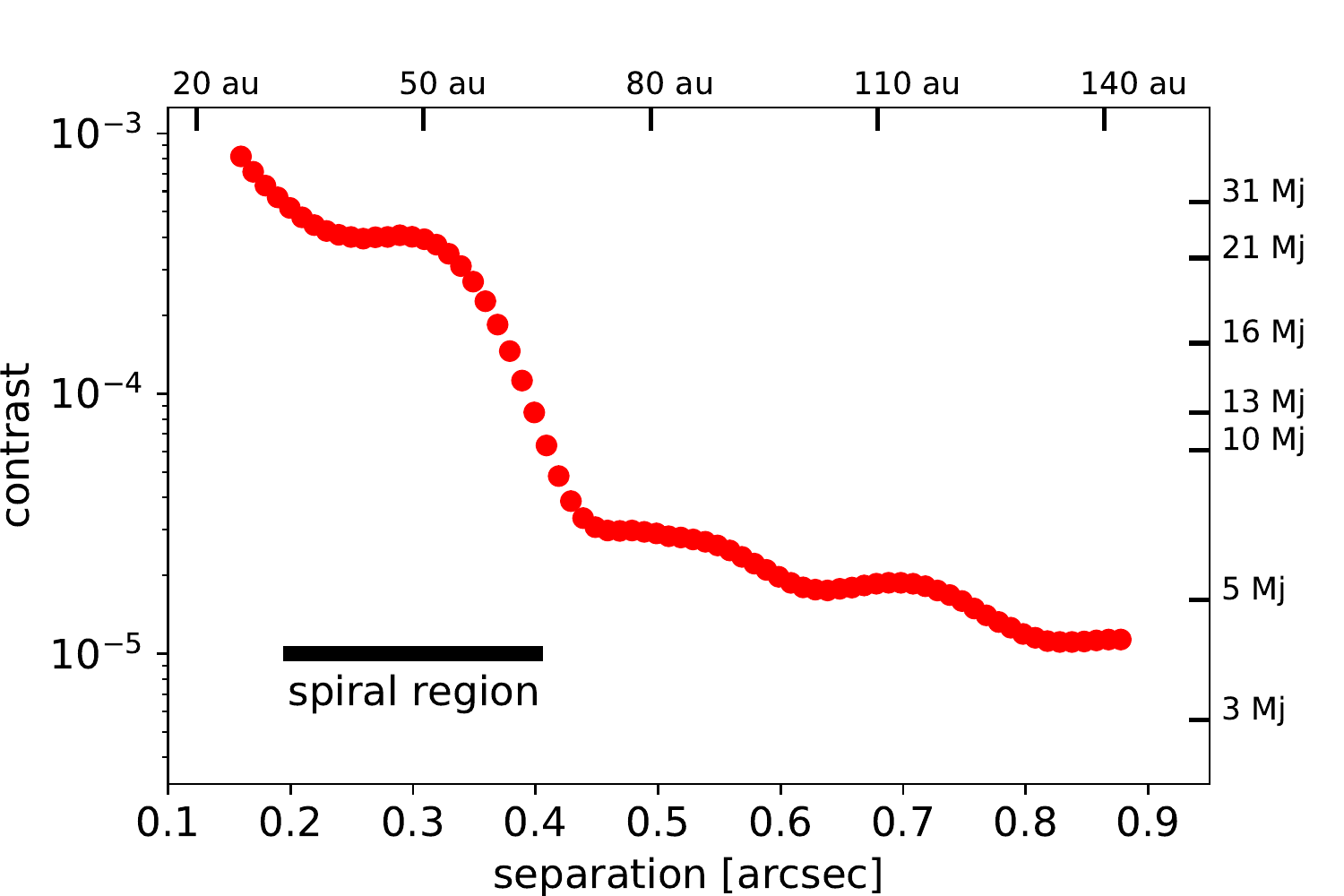}
    \caption{5$\sigma$ contrast limit of the NIRC2 image with PC=8. We also plot expected contrast of substellar-mass object on the right using the COND03 model.}
    \label{contrast NIRC2}
\end{figure}

Apart from the spiral feature, we did not detect any companion candidates within $\sim$$0\farcs9$ from the central star. The NIRC2 figure with a larger field of view (FoV) is shown in Appendix \ref{sec: Supplementary Keck/NIRC2 Images}. 
We then calculated noise profiles as a function of separation relative to the signal from the central star. Figure \ref{contrast NIRC2} shows a 5$\sigma$ detection limit of the NIRC2 data. Although the spiral feature affects the detection limit between $0\farcs2-0\farcs4$, we achieved 2.9$\times10^{-5}$ at 0$\farcs5$. Compared with an evolutionary model \citep[COND03;][]{Baraffe2003} assuming 10 Myr, our contrast limit could constrain down to $\sim$5 $M_{\rm J}$ outside the spiral.

\subsection{Subaru/HiCIAO} \label{sec: Subaru/HiCIAO}
\begin{figure*}
\begin{tabular}{cc}
    \begin{minipage}{0.4\hsize}
    \centering
    \includegraphics[scale=0.65]{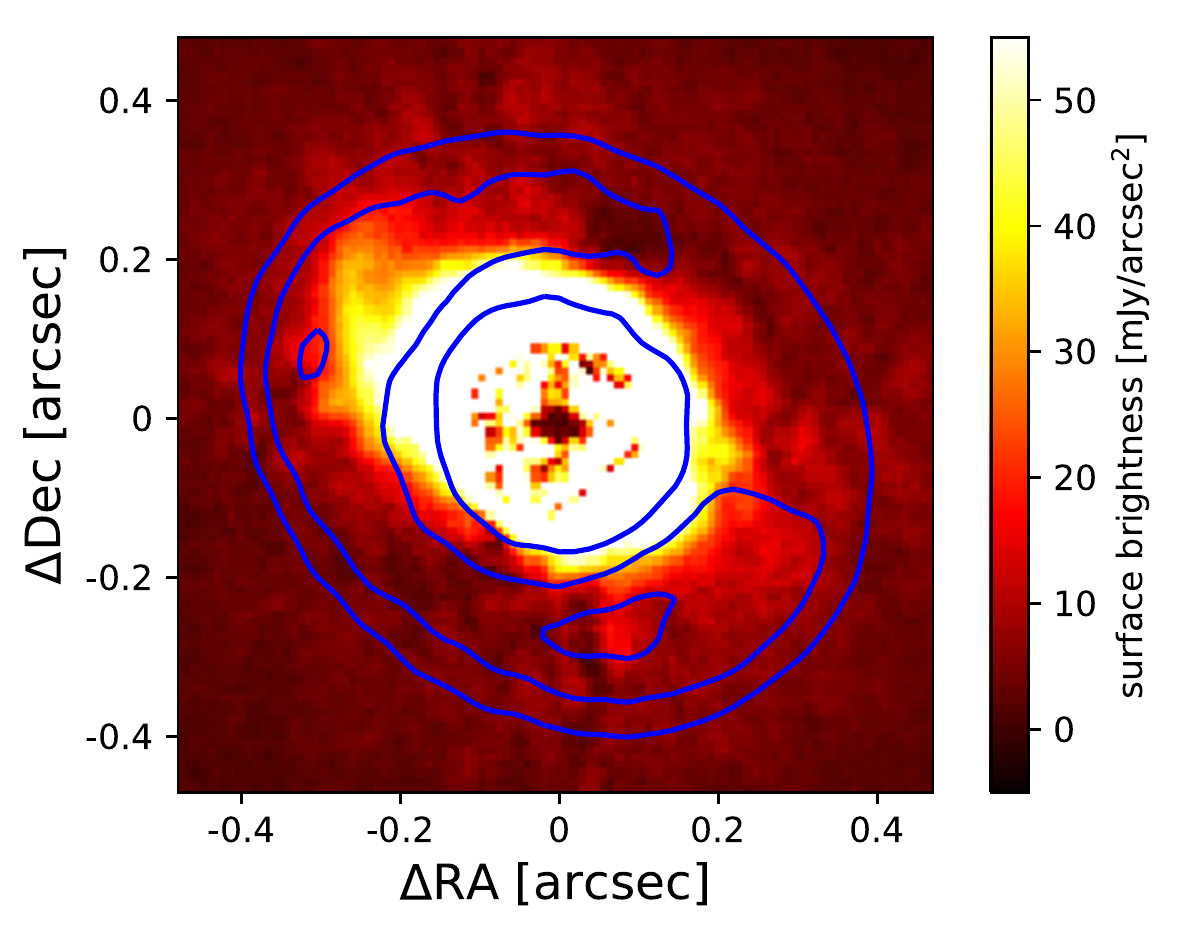}
    \end{minipage}
    \begin{minipage}{0.6\hsize}
    \centering
    \includegraphics[scale=0.23]{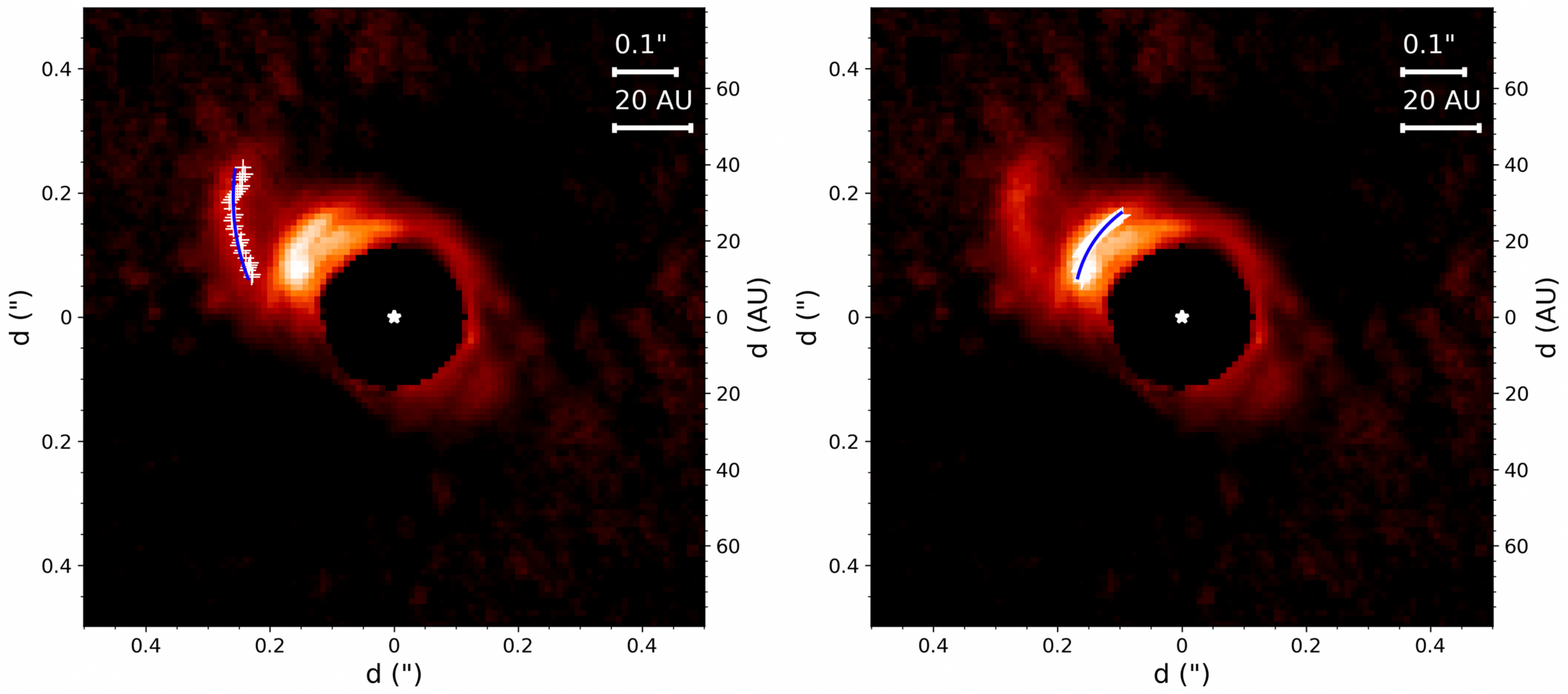}
    \end{minipage}
    \end{tabular}
    \caption{(left) PI image of Subaru/HiCIAO $H$-band observation overlaid with the ALMA continuum. North is up and east is left. (middle) $R^2$-scaled PI image including the identified trace of the outer spiral (white crosses) and the best-fit logarithmic spiral (blue curve). (right) Same for the inner (tentative) spiral. For estimating the pitch angle we used a further-deprojected image.}
    \label{fig: HiCIAO}
\end{figure*}

\begin{figure}
 \centering
 \includegraphics[width=0.48\textwidth]{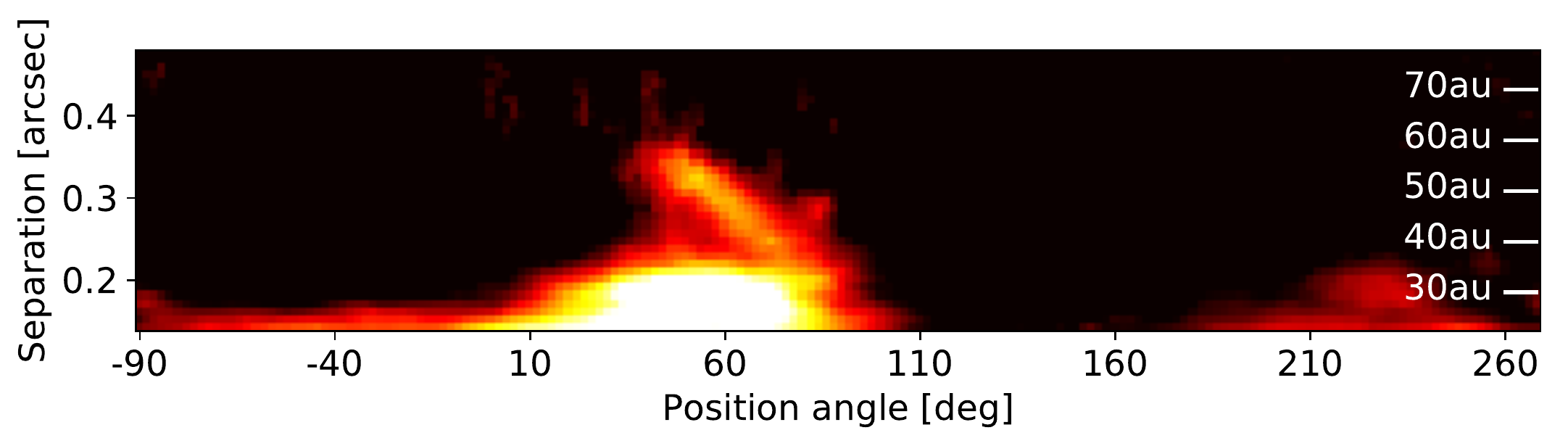}
 \caption{As Figure \ref{r-theta NIRC2} for the $r^2$-scaled HiCIAO PI image. }
 \label{r-theta HiCIAO}
\end{figure}

Subaru/AO188+HiCIAO observed CQ Tau in a combination of PDI and ADI as part of the SEEDS project \citep[][]{Tamura2009}.
No coronagraph was used in this observation.
The total exposure time of the HiCIAO data is only 9 min with FWHM=5.3 pix ($\sim$50 mas with a pixel scale of 9.5 mas/pix), which achieved an inner working angle of $\sim0\farcs77$ after the ADI reduction and is insufficient for searching planets embedded in the CQ Tau disk \citep[for the ADI result at separations $\geq1\farcs0$, see][]{Uyama2017}. In this study, we focus only on the PDI reduction. 
SEEDS adopted standard PDI (sPDI) and quad PDI (qPDI), where a different number of Wollaston prisms was used, and sPDI was applied to CQ Tau's observation \citep[for detailed information see][]{Uyama2017}.
After the first reduction of destriping the HiCIAO pattern, flat fielding, distortion correction, and image registration, we reduced the polarimetric data sets by means of an {\tt IRAF} pipeline\footnote{IRAF is
distributed by National Optical Astronomy Observatory, which is
operated by the Association of Universities for Research in Astronomy,
Inc., under cooperative agreement with the National Science
Foundation.}, which was used in previous HiCIAO PDI studies \citep[e.g.,][]{Hashimoto2011,Hashimoto2012}.
Figure \ref{fig: HiCIAO} shows the PI image of CQ Tau overlaid with ALMA continuum. 
The whole disk cannot be investigated since there are residual speckles that cannot be removed through post-processing due to short exposure time.
We did not detect a gap in the surface of CQ Tau's disk.
The PI image shows the spiral feature at the same location as shown in the NIRC2 image.
In order to investigate the SNR of the spiral, we used perpendicular regions to the spiral whose PAs range 125$^\circ$-165$^\circ$, 305$^\circ$-345$^\circ$ for calculating a noise (defined as standard deviation in the specified area) radial profile.
We finally confirmed that the spiral has an SNR$\sim$5--6 in the PI image. There may be other disk features shown in the PDI-reduced image but below 5$\sigma$ significance due to speckles in the inner region.
An $r^2$-scaled PI image (see Figure \ref{r-theta HiCIAO} for a polar-projected image) clearly shows the spiral feature.
There is another extended region at PAs between 10$^\circ$ and 90$^\circ$, which is perhaps another asymmetric feature and possibly detected in the NIRC2 data with PC=5, 8, and 10 in Figure \ref{fig: NIRC2 PCs} with insufficient significance.
We discuss this inner feature in Section \ref{sec: Commparison of the Two Data Sets}.
We note that a gap-like feature close to the central star may be affected by $r^2$-scaling because the original HiCIAO data set does not show such a feature (see the left image in Figure \ref{fig: HiCIAO} for the PI signal and the right image in Figure \ref{fig: NIRC2} for the contour).

\subsection{Comparison of the Two Data Sets} \label{sec: Commparison of the Two Data Sets}
In both observations we clearly detected the spiral feature, which overlaps with the ring 
structure in the millimeter continuum detected by ALMA. 
The presence of the spiral is consistent with a prediction of $\sim$6--9 $M_{\rm J}$ planet at 20 au \citep[][]{Ubeira2019}. 
However, our observation could not achieve a sufficient contrast limit to detect/constrain such a faint protoplanet.
We checked whether a counterpart of the spiral is shown in the ALMA gas data.
\cite{Tang2017} reported a pair of spirals for AB Aur at $^{12}$CO emission that correspond to the PI signal \citep[][]{Hashimoto2011}. However, \cite{Ubeira2019} did not show any clear spiral features in the $^{12}$CO data.

The right image in Figure \ref{fig: NIRC2} compares the NIRC2 and HiCIAO results and these shapes show a good agreement with each other.
Polar-projected images (Figures \ref{r-theta NIRC2} and \ref{r-theta HiCIAO}) also clearly show that the spiral feature increases in distance from the central star.
We note that the surface brightness in each band shows a different parameter. The NIRC2 and HiCIAO results correspond to total intensity and polarimetric intensity, respectively. We discuss the difference between these results in Section \ref{sec: scattering case}.
We measured the pitch angle based on the best-fit logarithmic spiral to the trace of the spiral. The trace was identified as radial maxima in azimuthal bins of $1\degr$ in the image obtained after deprojection using inclination and position angle of the major axis derived by \cite{Ubeira2019}: $i=35\degr$ and PA$_a = 55\degr$. 
Since the spiral feature in the NIRC2 image experiences self-subtraction and is distorted by the reduction algorithm, we used only the HiCIAO data to measure the pitch angle. The PI data corresponds to scattering profiles from the disk surface and does not experience self-subtraction.
The fitted result for the spiral (the central image in Figure \ref{fig: HiCIAO}) is $34\degr\pm2\degr$.
We also attempted to fit the extended inner region at PAs between 10$^\circ$ and 90$^\circ$. The result is shown in the right image of Figure \ref{fig: HiCIAO} and the pitch angle is measured at $4\degr\pm3\degr$ deg.
In addition to fits to the logarithmic spiral equation, we also fitted the spiral trace to the general Archimedean equation. The result is shown in Appendix \ref{sec: Supplementary Keck/NIRC2 Images}.
We note that because scattered light originate from a cone-shape surface instead of a flat plane, when viewed at a finite inclination, different regions in the disk are compressed differently (e.g., Figure 4, \citealt{ginski16}). Because of this, a disk structure in surface density traced by mm continuum emission can be projected to a different location in scattered light (e.g., the southern spiral arm in MWC 758, \citealt{dong18mwc758}). Simple deprojection by linearly expanding the disk along the minor axis by a factor of $1/\cos{i}$ generally does not perfectly restore the face-on view of the disk \citep{dong16armviewing}. Therefore, our measurements of the arm pitch angles are approximations only. Future modeling work is needed to simultaneously determine the shapes of the disk surface and the spiral arms.

\begin{figure*}
\begin{tabular}{cc}
\begin{minipage}{0.5\hsize}
    \centering
    \includegraphics[scale=0.65]{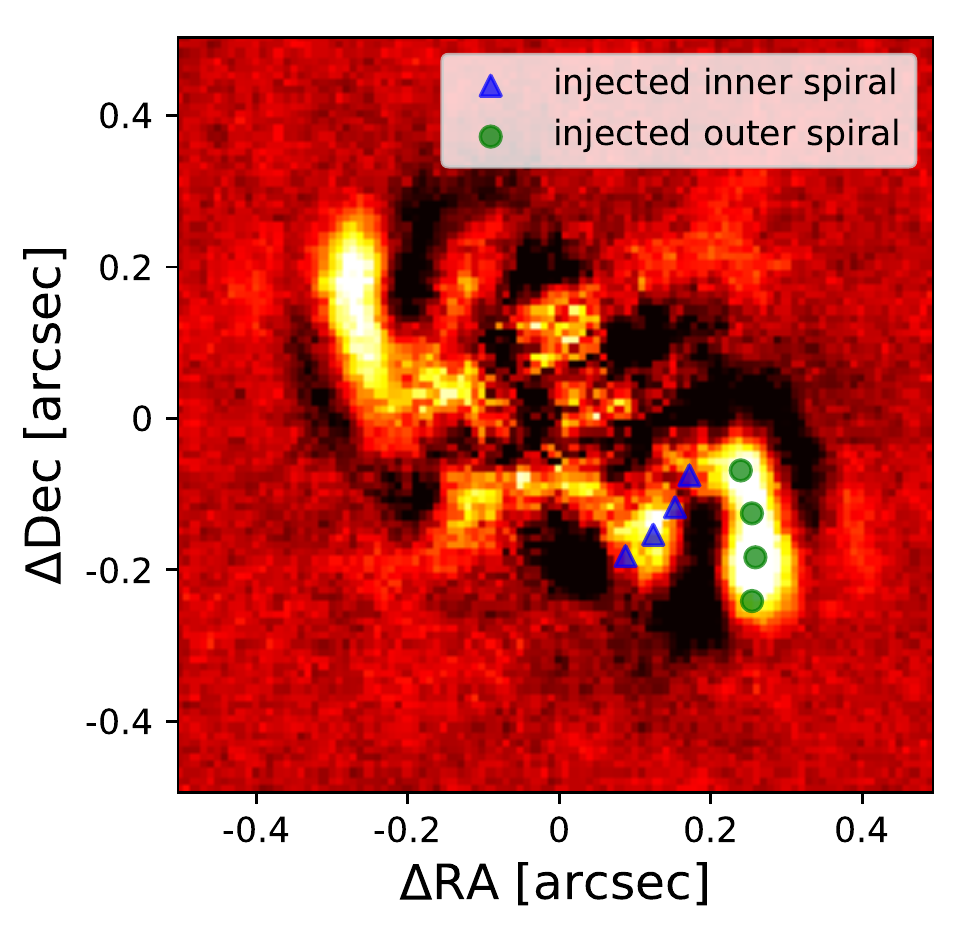}
\end{minipage}
\begin{minipage}{0.5\hsize}
    \centering
    \includegraphics[width=0.95\textwidth]{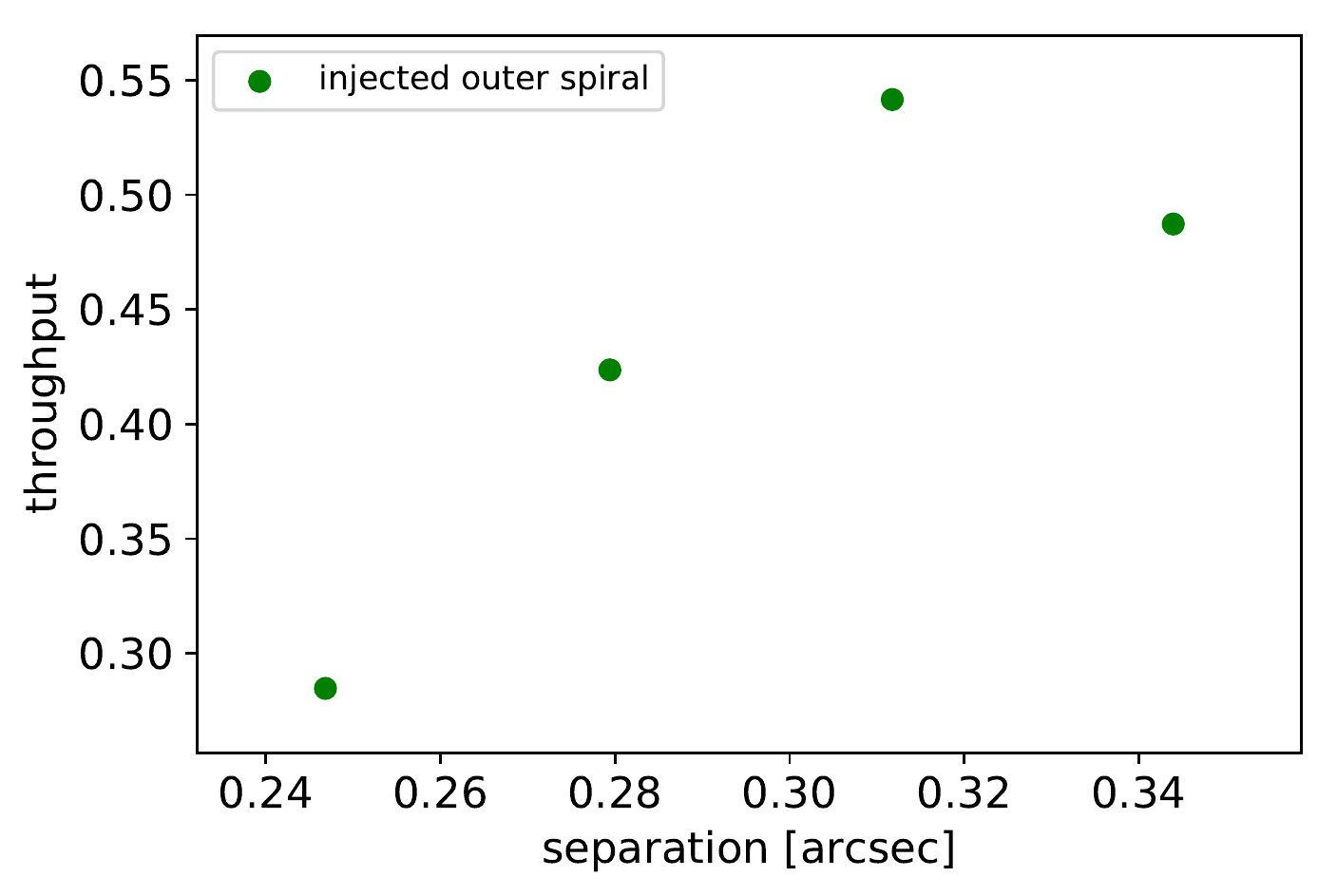}
\end{minipage}
\end{tabular}
    \caption{(left) VIP-PCA reduced image for the NIRC2 data with injected fake spirals at symmetric positions with respect to the center. (right) Measured throughput of the injected outer spiral as a function of separation.}
    \label{fig: forward modeling}
\end{figure*}

We used both fit results to infer input parameters for the forward modeling of the $L^\prime$-band feature \citep[for the detailed method for the forward modeling, see][]{Christiaens2019} to measure a throughput (signal loss due to the ADI reduction); Figure \ref{fig: forward modeling} shows our result with injected spirals.
We used the off-axis PSF of CQ Tau and injected fake PSFs at several separations and position angles to produce fake spiral features (injected positions shown in Figure \ref{fig: forward modeling}).
We then measured the ratio between input flux and output flux at the injected locations, which is shown in the right image of Figure \ref{fig: forward modeling}.
Our forward modeling reproduced the outer spiral with a flux recovered by a throughput of 0.54 at 50 au, which corresponds to 126 mJy/arcsec$^2$ at the brightest region in the spiral.
On the other hand, however, the injected inner spiral is largely affected (a throughput less than 0.3) by not only self-subtraction at small separations but also negative regions produced by the existence of the outer spiral.
As the SNRs of this feature in the practical NIRC2 and HiCIAO data are less than 5, we do not conclude that this inner feature is a spiral.

The CQ Tau disk has a striking similarity with the disk around V1247 --- both of which show one prominent arm in scattered light and a ringed disk in mm continuum emission \citep{Ohta16, kraus17}. In addition, they share a similar inclination of $\sim30-35^\circ$, and the major spiral arm seen is in the direction of the major axis. Simulations have shown that while a massive companion may induce a pair of nearly symmetric spiral arms, when viewed at a modest inclination one of the arms may be compressed more than the other in scattered light, thus falling inside the inner working angle \citep{dong16armviewing}. Future observations may push for inner separations to look for possible additional arms hidden under the current image mask.

\section{Discussion}
\label{sec: Discussion}
The spiral feature in $L^\prime$-band may be reproduced by two scenarios; 
1) the spiral is heated and self-luminous or 2) the surface of the spiral scatters the stellar light as shown in the Subaru/HiCIAO image.
Hereafter we use the peak values at the spiral region as a representative surface brightness.
The spiral features extends over $\sim0\farcs2$-scale, which is only a factor of $\sim2$ larger than the angular resolution of the NIRC2 result. This prevents us to discuss discuss the detailed spiral profiles of the surface brightness distribution.

\subsection{Disk Temperature} \label{sec: Hot Spiral Case}
We first investigate whether the L$^{\prime}$-band emission can be reproduced by thermal emission from small grains. We assume that the disk is optically thick at L$^\prime$-band.
If the observed emission of 126~mJy/arcsec$^2$ is entirely due to thermal emission from optically thick dust, the temperature of the grains is expected to be $T_{\rm grain}\sim$202.5K.

Provided that small grains at the spiral absorb shorter wavelengths of stellar light and emit their heat at $\sim$3--4 $\mu$m, the grain temperature is given by the following equation
\begin{eqnarray} \label{eq: grain temperature}
\begin{split}
&&(1-\omega_\nu)\pi a_{\rm grain}^2\frac{L_{\star}}{r_{\rm spiral}^2}<Q_{\rm abs}(\lambda_\star)> \\
&&=\ 4\pi a_{\rm grain}^2\sigma T_{\rm grain}^4Q_{\rm abs}(\lambda_{\rm grain}),
\end{split}
\end{eqnarray}
where $Q_{\rm abs}(\lambda)$ is the absorption efficiency at $\lambda$ and $a_{\rm grain}$ is the size of the grain. With this equation, we can derive a set of dust size ($a_{\rm grain}$) and albedo ($\omega_\nu)$ that can reproduce $T_{\rm grain}=202.5$K.
Here we assume $<Q_{\rm abs}(\lambda_\star)>=\frac{\int Q_{\rm abs}B_\lambda(T_\star) d\lambda}{\int B_\lambda(T_\star) d\lambda}\simeq1$ for simplicity. This relation is compared with the model of astronomical silicate \citep{Draine1984}. 
We find a set of $a_{\rm grain}\sim0.8\ \mu$m and albedo$\sim$0.2 can reproduce $T_{\rm grain}=202.5$ K and is consistent with the astronomical silicate model.
The dust opacity ($\kappa$) per unit dust mass (assuming gas-to-dust ratio is 100) is also estimated to be 3.4$\times10^3$~cm$^2$/g and 1.1$\times10^3$~cm$^2$/g at $H$- and $L^\prime$-bands, respectively.
With the small grain surface density of 0.0375~g/cm$^2$ of CQ Tau's disk \citep[][]{Ubeira2019}, $\tau_\nu$ is assessed at larger than 30, which corresponds to the optically-thick disk.
We note that current ALMA observations reported lower $\kappa$ per unit dust mass \citep[$\sim$100~cm$^2$/g at near IR;][]{Birnstiel2018-DSHARP}. With this value $\tau_\nu$ is estimated 3.75 and enough for the optically-thick disk.

With the temperature of 202.5~K, the disk aspect ratio ($H/r_{\rm spiral}$) at the location of the spiral structure is estimated to be 0.15. Here, $H=c_s/\Omega_{\rm K}$ is the disk scale height, where $c_s$ and $\Omega_{\rm K}$ are the sound speed and the Keplerian angular velocity, respectively.

\subsection{Scattering} \label{sec: scattering case}

We then investigate whether both of the results taken by Subaru/HiCIAO and Keck/NIRC2 can be reproduced by only scattering.
As a rough estimate of the surface brightness of the scattered light, we use Equation (9) in \citet{Inoue2008}, which is an approximate analytic expression of the scattered light \citep{Dalessio1999,DAlessio2006}.  With the modification for inclined disks \citep[see also][]{Jang-Condell2013}, 
the observable intensity is given by the following equation
\begin{eqnarray}\label{eq: scattering}
I^{\rm sca}_{\nu}\simeq\beta\omega_{\nu}H(1,\omega_\nu)B_\nu(T_\star)\frac{\Omega_\star}{4\pi}\frac{1}{\sin\beta+\cos\eta} ,
\end{eqnarray} 
where $\beta$, $\omega_\nu$, $\Omega_\star= \pi\frac{R_\star^2}{r_{\rm spiral}^2}$, and $T_\star$ are a grazing angle, albedo, a solid angle of the stellar photosphere from the spiral, and effective temperature of CQ Tau, respectively. $\eta$ is defined as the sum of inclination and $\arctan(\frac{dH(r)}{dr})\,\sim\,H/r$. 
$H(1,\omega_\nu)$ represents the law of diffuse reflection \citep{Chandrasekhar1960} and $B_\nu(T_\star)$ is Planck's law.
Here, we assume $H(1,\omega_\nu)=1$ (single scattering) for simplicity.\footnote{For a multi-scattering case, see also Equation (9) in \citet{Jang-Condell2013}.}

To estimate the surface brightness of the scattered light from the spiral region, we used the disk and dust parameters estimated in Section \ref{sec: Hot Spiral Case}.  With this dust model, the albedo is about 0.4 at $H$-band. 
The grazing angle ($\beta = dH/dr - \arctan({H/r})=f\times H/r$, where $f$ is a flaring index defined by $H(r)\propto r^{(1+f)}$) is assumed to be $\beta \lesssim H/r=0.15$ ($f\lesssim1$).
$R_\star$ was derived from Stefan-Boltzmann law with $T_\star$=6900 K and $L_\star=10L_{\odot}$ \citep[][]{Testi2001,Ubeira2019}. We used 50 au as a typical value for $r_{\rm spiral}$ and $i=35^\circ$ for the disk's inclination \citep[][]{Ubeira2019}.
We finally assessed the expected scattered brightness as 62 and 8.5 mJy/arcsec$^2$ at $H$- and $L^\prime$-band, respectively.
Note that we do not take into account of polarization degree in the $H$-band calculation.
In the HiCIAO result, the spiral has surface brightness of $\sim$30--40 mJy/arcsec$^2$ and does not show a large disagreement with the estimate. 
The NIRC2 result, however, is much brighter than the expected brightness in $L^\prime$-band.

We note that CQ Tau has IR excess in its SED \citep[][]{Ubeira2019}.  The inner disk with effective temperature of $\sim$1000 K may behave as another source of heating and scattering mechanisms at $L^\prime$-band.
Assuming that the inner disk contributes as the light source of the scattering more than the central star by an order of magnitude, the expected $L^\prime$-band brightness of the spiral is $\sim90$ mJy/arcsec$^2$ and is comparable to the NIRC2 result.
Therefore, both thermal emission and scattering may equally contribute to the spiral feature at $L^\prime$-band, if the disk is heated up to $\sim$200~K at $\sim$50~au and the inner part of the disk contributes as a light source of scattering.
Detailed discussions using radiative transfer modeling will help to have better understandings of the spiral feature.

\subsection{Formation Scenario of the Spiral}

We have seen that the disk temperature at $\sim$ 50~au can be $\sim$200~K, indicating that the disk aspect ratio is $\sim$0.15. The pitch angle of the spiral feature (in radian) is comparable with the disk aspect ratio if the spiral feature is due to spiral density wave in a differentially rotating disk \citep[e.g.,][]{Rafikov2002,Bae2018}.
However, the fitted pitch angle of the spiral from the $r^2$-scaled HiCIAO image ($34^\circ\pm2^\circ$) is much larger than the expected $H/r_{\rm spiral}=0.15$.
\citet{Tang2017} reported a large pitch angle in the AB Aur disk, which is similar to the CQ Tau case, and they predicted an unseen gaseous planet that coincides with the large pitch angle.
As mentioned in Section \ref{sec: Introduction}, a high-mass planet can induce shocks and heats the spiral enough to be detected in $L^\prime$-band \citep{Lyra2016}.
The heating of the spiral arms driven by a massive companion may occur due to shock heating \citep[$\leq$15-20 deg;][]{Zhu2015,Dong2015planet}.
In this sense the prediction of \citet{Ubeira2019} of an unseen protoplanet is consistent with a heated spiral scenario.

Several other mechanisms can induce a large pitch angle: gravitational instability \citep[$\leq$15--20 deg;][]{Dong2015GI} or shadow casting \citep[$\leq$20--25 deg;][]{Montesinos2016,Montesinos2018}
Note that these studies, in many cases, assume vertically isothermal disk temperature profile.
\citet{Juhasz2018} performed another simulation by assuming the disk surface is hotter than the midplane and showed that the spiral pitch angle near the surface can be more open compared to that at disk midplane.
As we mentioned in Section \ref{sec: Subaru/HiCIAO}, our fitted result of the pitch angle can be distorted by the inclination effect and we do not identify the mechanism to make a wide-open spiral.
Combining gas observations of different emission lines enables to estimate the vertical temperature profile of the disk \citep[][]{Akiyama2011,Akiyama2013} and such future observations will help to understand the thermal structure of the spiral.
Since the NIRC2 observation did not detect any companion candidates, follow-up observations to search for planets within 30 AU are required to further investigate this scenario.

\section{Conclusion}

We have newly detected a spiral in the CQ Tau disk using the Keck/NIRC2 $L^prime$-band imaging and Subaru/AO188+HiCIAO $H$-band polarimetric imaging observations.
The spiral feature overlaps with the ring structure in the ALMA continuum and is consistent with a prediction of jovian protoplanet at 20 au \citep{Ubeira2019}.
However, our observations did not confirm the presence of the gap reported by the ALMA observations. We did not detect any companion candidates within $0\farcs9$ from the central star either.
The 5$\sigma$ contrast limit could constrain down to $\sim$5 $M
_{\rm J}$ though it is affected by the spiral structures at separations of $0\farcs2-0\farcs4$.
We traced peaks of the spiral in the $r^2$-scaled HiCIAO image to derive a pitch angle of the spiral ($34^\circ\pm2^\circ$).
This fitted result is also used for forward modeling to reproduce the ADI-reduced NIRC2 image and we estimated the original surface brightness in $L^\prime$-band to be $\sim$126 mJy/arcsec$^2$ at most.

We have investigated whether the $L^{\prime}$-band emission can be reproduced by thermal emission. The observed emission of 126 mJy/arcsec$^2$ corresponds to the brightness temperature temperature of $\sim$202.5 K.
The temperature of dust grainds at the spiral location can be $\sim200$ K if the grain size is $\sim$0.8 $\mu$m and the albedo is $\sim$0.2 for an astronomical silicate model \citep{Draine1984}.
The $H$-band emission originates from the scattering from the disk surface while both scattering and thermal emission may contribute to the $L^\prime$-band emission, depending on the condition of the inner disk.

Follow-up observations for the disk exploration as well as the companion search will help to understand this bright spiral feature.
The $L^{\prime}$-band profile for the spiral feature may be improved by high-contrast imaging with RDI.
PDI observations with an updated AO instrument such as Subaru/SCExAO, Gemini/GPI, or VLT/SPHERE will also be able to provide clearer images of the disk, which helps to understand the disk characteristics by spatially resolving the spiral.
Combining gas observations of different emission lines enables to estimate the vertical temperature profile of the disk.

\acknowledgments
The authors would like to thank the anonymous referees for their constructive comments and suggestions to improve the quality of the paper.
We wish to thank Mitsuhiko Honda for constructive comments to improve our discussions with the NIRC2 result.
Some of the data presented herein were obtained at the W. M. Keck Observatory, which is operated as a scientific partnership among the California Institute of Technology, the University of California and the National Aeronautics and Space Administration. The Observatory was made possible by the generous financial support of the W. M. Keck Foundation.
A part of this research is based on data collected at the Subaru Telescope, which is operated by the National Astronomical Observatories of Japan.
Based in part on data collected at Subaru telescope and obtained from the SMOKA, which is operated by the Astronomy Data Center, National Astronomical Observatory of Japan.
This paper makes use of the following ALMA data: ADS/JAO.ALMA\#2017.1.01404.S. ALMA is a partnership of ESO (representing its member states), NSF (USA) and NINS (Japan), together with NRC (Canada), MOST and ASIAA (Taiwan), and KASI (Republic of Korea), in cooperation with the Republic of Chile. The Joint ALMA Observatory is operated by ESO, AUI/NRAO and NAOJ.
This research has made use of NASA's Astrophysics Data System Bibliographic Services.
This research has made use of the SIMBAD database, operated at CDS, Strasbourg, France. 
This research has made use of the VizieR catalogue access tool, CDS, Strasbourg, France. The original description of the VizieR service was published in A{\&}AS 143, 23.

TU acknowledges JSPS overseas research fellowship.
This work was supported by MEXT/JSPS KAKENHI Grant Numbers 15H02063, 17K05399, 18H05442, 19H00703, 19H05089, and 19K03932.
Part of this work was carried out at the Jet Propulsion Laboratory, California Institute of Technology, under contract with the National Aeronautics and Space Administration (NASA).
The material is based upon work supported by NASA under award No 80NSSC19K0294.
The research leading to these results has received funding from the European Research Council under the European Union's Seventh Framework Program (ERC Grant Agreement n. 337569) and Horizon 2020 Research and Innovation Program (grant agreement No 819155), and from the Wallonia-Brussels Federation (grant for Concerted Research Actions).
VC acknowledges funding from the Australian Research Council via DP180104235.
OA acknowledges funding from FRS-FNRS. JB acknowledges support by NASA through the NASA Hubble Fellowship grant \#HST-HF2-51427.001-A awarded  by  the  Space  Telescope  Science  Institute,  which  is  operated  by  the  Association  of  Universities  for  Research  in  Astronomy, Incorporated, under NASA contract NAS5-26555.


The authors wish to acknowledge the very significant cultural role and reverence that the summit of Mauna Kea has always had within the indigenous Hawaiian community. We are most fortunate to have the opportunity to conduct observations from this mountain.

\bibliographystyle{aasjournal}                                                              
\bibliography{library}                                                                

\begin{thebibliography}{}
\expandafter\ifx\csname natexlab\endcsname\relax\def\natexlab#1{#1}\fi
\providecommand{\url}[1]{\href{#1}{#1}}
\providecommand{\dodoi}[1]{doi:~\href{http://doi.org/#1}{\nolinkurl{#1}}}
\providecommand{\doeprint}[1]{\href{http://ascl.net/#1}{\nolinkurl{http://ascl.net/#1}}}
\providecommand{\doarXiv}[1]{\href{https://arxiv.org/abs/#1}{\nolinkurl{https://arxiv.org/abs/#1}}}

\bibitem[{{Akiyama} {et~al.}(2011){Akiyama}, {Momose}, {Hayashi}, \&
  {Kitamura}}]{Akiyama2011}
{Akiyama}, E., {Momose}, M., {Hayashi}, H., \& {Kitamura}, Y. 2011, \pasj, 63,
  1059, \dodoi{10.1093/pasj/63.5.1059}

\bibitem[{{Akiyama} {et~al.}(2013){Akiyama}, {Momose}, {Kitamura},
  {Tsukagoshi}, {Shimada}, {Koyamatsu}, \& {Hayashi}}]{Akiyama2013}
{Akiyama}, E., {Momose}, M., {Kitamura}, Y., {et~al.} 2013, \pasj, 65, 123,
  \dodoi{10.1093/pasj/65.6.123}

\bibitem[{{ALMA Partnership} {et~al.}(2015){ALMA Partnership}, {Brogan},
  {P{\'e}rez}, {Hunter}, {Dent}, {Hales}, {Hills}, {Corder}, {Fomalont},
  {Vlahakis}, {Asaki}, {Barkats}, {Hirota}, {Hodge}, {Impellizzeri}, {Kneissl},
  {Liuzzo}, {Lucas}, {Marcelino}, {Matsushita}, {Nakanishi}, {Phillips},
  {Richards}, {Toledo}, {Aladro}, {Broguiere}, {Cortes}, {Cortes}, {Espada},
  {Galarza}, {Garcia-Appadoo}, {Guzman-Ramirez}, {Humphreys}, {Jung}, {Kameno},
  {Laing}, {Leon}, {Marconi}, {Mignano}, {Nikolic}, {Nyman}, {Radiszcz},
  {Remijan}, {Rod{\'o}n}, {Sawada}, {Takahashi}, {Tilanus}, {Vila Vilaro},
  {Watson}, {Wiklind}, {Akiyama}, {Chapillon}, {de Gregorio-Monsalvo}, {Di
  Francesco}, {Gueth}, {Kawamura}, {Lee}, {Nguyen Luong}, {Mangum}, {Pietu},
  {Sanhueza}, {Saigo}, {Takakuwa}, {Ubach}, {van Kempen}, {Wootten},
  {Castro-Carrizo}, {Francke}, {Gallardo}, {Garcia}, {Gonzalez}, {Hill},
  {Kaminski}, {Kurono}, {Liu}, {Lopez}, {Morales}, {Plarre}, {Schieven},
  {Testi}, {Videla}, {Villard}, {Andreani}, {Hibbard}, \&
  {Tatematsu}}]{ALMA2015}
{ALMA Partnership}, {Brogan}, C.~L., {P{\'e}rez}, L.~M., {et~al.} 2015, \apjl,
  808, L3, \dodoi{10.1088/2041-8205/808/1/L3}

\bibitem[{{Amara} \& {Quanz}(2012)}]{Amara2012}
{Amara}, A., \& {Quanz}, S.~P. 2012, Monthly Notices of the Royal Astronomical
  Society, 427, 948, \dodoi{10.1111/j.1365-2966.2012.21918.x}

\bibitem[{{Andrews} {et~al.}(2018){Andrews}, {Huang}, {P{\'e}rez}, {Isella},
  {Dullemond}, {Kurtovic}, {Guzm{\'a}n}, {Carpenter}, {Wilner}, {Zhang}, {Zhu},
  {Birnstiel}, {Bai}, {Benisty}, {Hughes}, {{\"O}berg}, \&
  {Ricci}}]{DSHARP-Andrews2018}
{Andrews}, S.~M., {Huang}, J., {P{\'e}rez}, L.~M., {et~al.} 2018, \apjl, 869,
  L41, \dodoi{10.3847/2041-8213/aaf741}

\bibitem[{{Bae} \& {Zhu}(2018)}]{Bae2018}
{Bae}, J., \& {Zhu}, Z. 2018, \apj, 859, 119, \dodoi{10.3847/1538-4357/aabf93}

\bibitem[{{Baraffe} {et~al.}(2003){Baraffe}, {Chabrier}, {Barman}, {Allard}, \&
  {Hauschildt}}]{Baraffe2003}
{Baraffe}, I., {Chabrier}, G., {Barman}, T.~S., {Allard}, F., \& {Hauschildt},
  P.~H. 2003, \aap, 402, 701, \dodoi{10.1051/0004-6361:20030252}

\bibitem[{{Benisty} {et~al.}(2015){Benisty}, {Juhasz}, {Boccaletti},
  {Avenhaus}, {Milli}, {Thalmann}, {Dominik}, {Pinilla}, {Buenzli}, {Pohl},
  {Beuzit}, {Birnstiel}, {de Boer}, {Bonnefoy}, {Chauvin}, {Christiaens},
  {Garufi}, {Grady}, {Henning}, {Huelamo}, {Isella}, {Langlois}, {M{\'e}nard},
  {Mouillet}, {Olofsson}, {Pantin}, {Pinte}, \& {Pueyo}}]{Benisty2015}
{Benisty}, M., {Juhasz}, A., {Boccaletti}, A., {et~al.} 2015, \aap, 578, L6,
  \dodoi{10.1051/0004-6361/201526011}

\bibitem[{{Birnstiel} {et~al.}(2018){Birnstiel}, {Dullemond}, {Zhu}, {Andrews},
  {Bai}, {Wilner}, {Carpenter}, {Huang}, {Isella}, {Benisty}, {P{\'e}rez}, \&
  {Zhang}}]{Birnstiel2018-DSHARP}
{Birnstiel}, T., {Dullemond}, C.~P., {Zhu}, Z., {et~al.} 2018, \apjl, 869, L45,
  \dodoi{10.3847/2041-8213/aaf743}

\bibitem[{{Chandrasekhar}(1960)}]{Chandrasekhar1960}
{Chandrasekhar}, S. 1960, {Radiative transfer}

\bibitem[{{Chapillon} {et~al.}(2010){Chapillon}, {Parise}, {Guilloteau},
  {Dutrey}, \& {Wakelam}}]{Chapillon2010}
{Chapillon}, E., {Parise}, B., {Guilloteau}, S., {Dutrey}, A., \& {Wakelam}, V.
  2010, \aap, 520, A61, \dodoi{10.1051/0004-6361/201014841}

\bibitem[{{Christiaens} {et~al.}(2019){Christiaens}, {Casassus}, {Absil},
  {Cantalloube}, {Gomez Gonzalez}, {Girard}, {Ram{\'\i}rez}, {Pairet},
  {Salinas}, {Price}, {Pinte}, {Quanz}, {Jord{\'a}n}, {Mawet}, \&
  {Wahhaj}}]{Christiaens2019}
{Christiaens}, V., {Casassus}, S., {Absil}, O., {et~al.} 2019, \mnras, 486,
  5819, \dodoi{10.1093/mnras/stz1232}

\bibitem[{{Currie} {et~al.}(2015){Currie}, {Cloutier}, {Brittain}, {Grady},
  {Burrows}, {Muto}, {Kenyon}, \& {Kuchner}}]{2015ApJ...814L..27C}
{Currie}, T., {Cloutier}, R., {Brittain}, S., {et~al.} 2015, \apjl, 814, L27,
  \dodoi{10.1088/2041-8205/814/2/L27}

\bibitem[{{D'Alessio} {et~al.}(2006){D'Alessio}, {Calvet}, {Hartmann},
  {Franco-Hern{\'a}ndez}, \& {Serv{\'\i}n}}]{DAlessio2006}
{D'Alessio}, P., {Calvet}, N., {Hartmann}, L., {Franco-Hern{\'a}ndez}, R., \&
  {Serv{\'\i}n}, H. 2006, \apj, 638, 314, \dodoi{10.1086/498861}

\bibitem[{{D'Alessio} {et~al.}(1999){D'Alessio}, {Calvet}, {Hartmann},
  {Lizano}, \& {Cant{\'o}}}]{Dalessio1999}
{D'Alessio}, P., {Calvet}, N., {Hartmann}, L., {Lizano}, S., \& {Cant{\'o}}, J.
  1999, \apj, 527, 893, \dodoi{10.1086/308103}

\bibitem[{{Dong} {et~al.}(2016){Dong}, {Fung}, \& {Chiang}}]{dong16armviewing}
{Dong}, R., {Fung}, J., \& {Chiang}, E. 2016, \apj, 826, 75,
  \dodoi{10.3847/0004-637X/826/1/75}

\bibitem[{{Dong} {et~al.}(2015{\natexlab{a}}){Dong}, {Hall}, {Rice}, \&
  {Chiang}}]{Dong2015GI}
{Dong}, R., {Hall}, C., {Rice}, K., \& {Chiang}, E. 2015{\natexlab{a}}, \apjl,
  812, L32, \dodoi{10.1088/2041-8205/812/2/L32}

\bibitem[{{Dong} {et~al.}(2018{\natexlab{a}}){Dong}, {Najita}, \&
  {Brittain}}]{Dong2018}
{Dong}, R., {Najita}, J.~R., \& {Brittain}, S. 2018{\natexlab{a}}, \apj, 862,
  103, \dodoi{10.3847/1538-4357/aaccfc}

\bibitem[{{Dong} {et~al.}(2015{\natexlab{b}}){Dong}, {Zhu}, {Rafikov}, \&
  {Stone}}]{Dong2015planet}
{Dong}, R., {Zhu}, Z., {Rafikov}, R.~R., \& {Stone}, J.~M. 2015{\natexlab{b}},
  \apjl, 809, L5, \dodoi{10.1088/2041-8205/809/1/L5}

\bibitem[{{Dong} {et~al.}(2018{\natexlab{b}}){Dong}, {Liu}, {Eisner},
  {Andrews}, {Fung}, {Zhu}, {Chiang}, {Hashimoto}, {Liu}, {Casassus},
  {Esposito}, {Hasegawa}, {Muto}, {Pavlyuchenkov}, {Wilner}, {Akiyama},
  {Tamura}, \& {Wisniewski}}]{dong18mwc758}
{Dong}, R., {Liu}, S.-y., {Eisner}, J., {et~al.} 2018{\natexlab{b}}, \apj, 860,
  124, \dodoi{10.3847/1538-4357/aac6cb}

\bibitem[{{Draine} \& {Lee}(1984)}]{Draine1984}
{Draine}, B.~T., \& {Lee}, H.~M. 1984, \apj, 285, 89, \dodoi{10.1086/162480}

\bibitem[{{Gaia Collaboration} {et~al.}(2018){Gaia Collaboration}, {Brown},
  {Vallenari}, {Prusti}, {de Bruijne}, {Babusiaux}, {Bailer-Jones}, {Biermann},
  {Evans}, {Eyer}, {Jansen}, {Jordi}, {Klioner}, {Lammers}, {Lindegren},
  {Luri}, {Mignard}, {Panem}, {Pourbaix}, {Randich}, {Sartoretti}, {Siddiqui},
  {Soubiran}, {van Leeuwen}, {Walton}, {Arenou}, {Bastian}, {Cropper},
  {Drimmel}, {Katz}, {Lattanzi}, {Bakker}, {Cacciari}, {Casta{\~n}eda},
  {Chaoul}, {Cheek}, {De Angeli}, {Fabricius}, {Guerra}, {Holl}, {Masana},
  {Messineo}, {Mowlavi}, {Nienartowicz}, {Panuzzo}, {Portell}, {Riello},
  {Seabroke}, {Tanga}, {Th{\'e}venin}, {Gracia-Abril}, {Comoretto},
  {Garcia-Reinaldos}, {Teyssier}, {Altmann}, {Andrae}, {Audard},
  {Bellas-Velidis}, {Benson}, {Berthier}, {Blomme}, {Burgess}, {Busso},
  {Carry}, {Cellino}, {Clementini}, {Clotet}, {Creevey}, {Davidson}, {De
  Ridder}, {Delchambre}, {Dell'Oro}, {Ducourant},
  {Fern{\'a}ndez-Hern{\'a}ndez}, {Fouesneau}, {Fr{\'e}mat}, {Galluccio},
  {Garc{\'\i}a-Torres}, {Gonz{\'a}lez-N{\'u}{\~n}ez}, {Gonz{\'a}lez-Vidal},
  {Gosset}, {Guy}, {Halbwachs}, {Hambly}, {Harrison}, {Hern{\'a}ndez},
  {Hestroffer}, {Hodgkin}, {Hutton}, {Jasniewicz}, {Jean-Antoine-Piccolo},
  {Jordan}, {Korn}, {Krone-Martins}, {Lanzafame}, {Lebzelter}, {L{\"o}ffler},
  {Manteiga}, {Marrese}, {Mart{\'\i}n-Fleitas}, {Moitinho}, {Mora}, {Muinonen},
  {Osinde}, {Pancino}, {Pauwels}, {Petit}, {Recio-Blanco}, {Richards},
  {Rimoldini}, {Robin}, {Sarro}, {Siopis}, {Smith}, {Sozzetti}, {S{\"u}veges},
  {Torra}, {van Reeven}, {Abbas}, {Abreu Aramburu}, {Accart}, {Aerts},
  {Altavilla}, {{\'A}lvarez}, {Alvarez}, {Alves}, {Anderson}, {Andrei},
  {Anglada Varela}, {Antiche}, {Antoja}, {Arcay}, {Astraatmadja}, {Bach},
  {Baker}, {Balaguer-N{\'u}{\~n}ez}, {Balm}, {Barache}, {Barata}, {Barbato},
  {Barblan}, {Barklem}, {Barrado}, {Barros}, {Barstow}, {Bartholom{\'e}
  Mu{\~n}oz}, {Bassilana}, {Becciani}, {Bellazzini}, {Berihuete}, {Bertone},
  {Bianchi}, {Bienaym{\'e}}, {Blanco-Cuaresma}, {Boch}, {Boeche}, {Bombrun},
  {Borrachero}, {Bossini}, {Bouquillon}, {Bourda}, {Bragaglia}, {Bramante},
  {Breddels}, {Bressan}, {Brouillet}, {Br{\"u}semeister}, {Brugaletta},
  {Bucciarelli}, {Burlacu}, {Busonero}, {Butkevich}, {Buzzi}, {Caffau},
  {Cancelliere}, {Cannizzaro}, {Cantat-Gaudin}, {Carballo}, {Carlucci},
  {Carrasco}, {Casamiquela}, {Castellani}, {Castro-Ginard}, {Charlot},
  {Chemin}, {Chiavassa}, {Cocozza}, {Costigan}, {Cowell}, {Crifo}, {Crosta},
  {Crowley}, {Cuypers}, {Dafonte}, {Damerdji}, {Dapergolas}, {David}, {David},
  {de Laverny}, {De Luise}, {De March}, {de Martino}, {de Souza}, {de Torres},
  {Debosscher}, {del Pozo}, {Delbo}, {Delgado}, {Delgado}, {Di Matteo},
  {Diakite}, {Diener}, {Distefano}, {Dolding}, {Drazinos}, {Dur{\'a}n},
  {Edvardsson}, {Enke}, {Eriksson}, {Esquej}, {Eynard Bontemps}, {Fabre},
  {Fabrizio}, {Faigler}, {Falc{\~a}o}, {Farr{\`a}s Casas}, {Federici},
  {Fedorets}, {Fernique}, {Figueras}, {Filippi}, {Findeisen}, {Fonti},
  {Fraile}, {Fraser}, {Fr{\'e}zouls}, {Gai}, {Galleti}, {Garabato},
  {Garc{\'\i}a-Sedano}, {Garofalo}, {Garralda}, {Gavel}, {Gavras}, {Gerssen},
  {Geyer}, {Giacobbe}, {Gilmore}, {Girona}, {Giuffrida}, {Glass}, {Gomes},
  {Granvik}, {Gueguen}, {Guerrier}, {Guiraud}, {Guti{\'e}rrez-S{\'a}nchez},
  {Haigron}, {Hatzidimitriou}, {Hauser}, {Haywood}, {Heiter}, {Helmi}, {Heu},
  {Hilger}, {Hobbs}, {Hofmann}, {Holland}, {Huckle}, {Hypki}, {Icardi},
  {Jan{\ss}en}, {Jevardat de Fombelle}, {Jonker}, {Juh{\'a}sz}, {Julbe},
  {Karampelas}, {Kewley}, {Klar}, {Kochoska}, {Kohley}, {Kolenberg},
  {Kontizas}, {Kontizas}, {Koposov}, {Kordopatis}, {Kostrzewa-Rutkowska},
  {Koubsky}, {Lambert}, {Lanza}, {Lasne}, {Lavigne}, {Le Fustec}, {Le
  Poncin-Lafitte}, {Lebreton}, {Leccia}, {Leclerc}, {Lecoeur-Taibi},
  {Lenhardt}, {Leroux}, {Liao}, {Licata}, {Lindstr{\o}m}, {Lister}, {Livanou},
  {Lobel}, {L{\'o}pez}, {Managau}, {Mann}, {Mantelet}, {Marchal}, {Marchant},
  {Marconi}, {Marinoni}, {Marschalk{\'o}}, {Marshall}, {Martino}, {Marton},
  {Mary}, {Massari}, {Matijevi{\v{c}}}, {Mazeh}, {McMillan}, {Messina},
  {Michalik}, {Millar}, {Molina}, {Molinaro}, {Moln{\'a}r}, {Montegriffo},
  {Mor}, {Morbidelli}, {Morel}, {Morris}, {Mulone}, {Muraveva}, {Musella},
  {Nelemans}, {Nicastro}, {Noval}, {O'Mullane}, {Ord{\'e}novic},
  {Ord{\'o}{\~n}ez-Blanco}, {Osborne}, {Pagani}, {Pagano}, {Pailler},
  {Palacin}, {Palaversa}, {Panahi}, {Pawlak}, {Piersimoni}, {Pineau}, {Plachy},
  {Plum}, {Poggio}, {Poujoulet}, {Pr{\v{s}}a}, {Pulone}, {Racero}, {Ragaini},
  {Rambaux}, {Ramos-Lerate}, {Regibo}, {Reyl{\'e}}, {Riclet}, {Ripepi}, {Riva},
  {Rivard}, {Rixon}, {Roegiers}, {Roelens}, {Romero-G{\'o}mez}, {Rowell},
  {Royer}, {Ruiz-Dern}, {Sadowski}, {Sagrist{\`a} Sell{\'e}s}, {Sahlmann},
  {Salgado}, {Salguero}, {Sanna}, {Santana-Ros}, {Sarasso}, {Savietto},
  {Schultheis}, {Sciacca}, {Segol}, {Segovia}, {S{\'e}gransan}, {Shih},
  {Siltala}, {Silva}, {Smart}, {Smith}, {Solano}, {Solitro}, {Sordo}, {Soria
  Nieto}, {Souchay}, {Spagna}, {Spoto}, {Stampa}, {Steele},
  {Steidelm{\"u}ller}, {Stephenson}, {Stoev}, {Suess}, {Surdej}, {Szabados},
  {Szegedi-Elek}, {Tapiador}, {Taris}, {Tauran}, {Taylor}, {Teixeira},
  {Terrett}, {Teyssand ier}, {Thuillot}, {Titarenko}, {Torra Clotet}, {Turon},
  {Ulla}, {Utrilla}, {Uzzi}, {Vaillant}, {Valentini}, {Valette}, {van Elteren},
  {Van Hemelryck}, {van Leeuwen}, {Vaschetto}, {Vecchiato}, {Veljanoski},
  {Viala}, {Vicente}, {Vogt}, {von Essen}, {Voss}, {Votruba}, {Voutsinas},
  {Walmsley}, {Weiler}, {Wertz}, {Wevers}, {Wyrzykowski}, {Yoldas},
  {{\v{Z}}erjal}, {Ziaeepour}, {Zorec}, {Zschocke}, {Zucker}, {Zurbach}, \&
  {Zwitter}}]{Gaia2018DR2}
{Gaia Collaboration}, {Brown}, A.~G.~A., {Vallenari}, A., {et~al.} 2018, \aap,
  616, A1, \dodoi{10.1051/0004-6361/201833051}

\bibitem[{{Ginski} {et~al.}(2016){Ginski}, {Stolker}, {Pinilla}, {Dominik},
  {Boccaletti}, {de Boer}, {Benisty}, {Biller}, {Feldt}, {Garufi}, {Keller},
  {Kenworthy}, {Maire}, {M{\'e}nard}, {Mesa}, {Milli}, {Min}, {Pinte}, {Quanz},
  {van Boekel}, {Bonnefoy}, {Chauvin}, {Desidera}, {Gratton}, {Girard},
  {Keppler}, {Kopytova}, {Lagrange}, {Langlois}, {Rouan}, \&
  {Vigan}}]{ginski16}
{Ginski}, C., {Stolker}, T., {Pinilla}, P., {et~al.} 2016, \aap, 595, A112,
  \dodoi{10.1051/0004-6361/201629265}

\bibitem[{{Gomez Gonzalez} {et~al.}(2017){Gomez Gonzalez}, {Wertz}, {Absil},
  {Christiaens}, {Defr{\`e}re}, {Mawet}, {Milli}, {Absil}, {Van Droogenbroeck},
  \& {Cantalloube}}]{Gonzalez2017}
{Gomez Gonzalez}, C.~A., {Wertz}, O., {Absil}, O., {et~al.} 2017, \aj, 154, 7,
  \dodoi{10.3847/1538-3881/aa73d7}

\bibitem[{{Hashimoto} {et~al.}(2011){Hashimoto}, {Tamura}, {Muto}, {Kudo},
  {Fukagawa}, {Fukue}, {Goto}, {Grady}, {Henning}, \& {Hodapp}}]{Hashimoto2011}
{Hashimoto}, J., {Tamura}, M., {Muto}, T., {et~al.} 2011, \apjl, 729, L17,
  \dodoi{10.1088/2041-8205/729/2/L17}

\bibitem[{{Hashimoto} {et~al.}(2012){Hashimoto}, {Dong}, {Kudo}, {Honda},
  {McClure}, {Zhu}, {Muto}, {Wisniewski}, {Abe}, \& {Brandner}}]{Hashimoto2012}
{Hashimoto}, J., {Dong}, R., {Kudo}, T., {et~al.} 2012, \apjl, 758, L19,
  \dodoi{10.1088/2041-8205/758/1/L19}

\bibitem[{{Huang} {et~al.}(2018){Huang}, {Andrews}, {P{\'e}rez}, {Zhu},
  {Dullemond}, {Isella}, {Benisty}, {Bai}, {Birnstiel}, {Carpenter},
  {Guzm{\'a}n}, {Hughes}, {{\"O}berg}, {Ricci}, {Wilner}, \&
  {Zhang}}]{DSHARP-Huang2018}
{Huang}, J., {Andrews}, S.~M., {P{\'e}rez}, L.~M., {et~al.} 2018, \apjl, 869,
  L43, \dodoi{10.3847/2041-8213/aaf7a0}

\bibitem[{{Inoue} {et~al.}(2008){Inoue}, {Honda}, {Nakamoto}, \&
  {Oka}}]{Inoue2008}
{Inoue}, A.~K., {Honda}, M., {Nakamoto}, T., \& {Oka}, A. 2008, \pasj, 60, 557,
  \dodoi{10.1093/pasj/60.3.557}

\bibitem[{{Jang-Condell} \& {Turner}(2013)}]{Jang-Condell2013}
{Jang-Condell}, H., \& {Turner}, N.~J. 2013, \apj, 772, 34,
  \dodoi{10.1088/0004-637X/772/1/34}

\bibitem[{{Juh{\'a}sz} \& {Rosotti}(2018)}]{Juhasz2018}
{Juh{\'a}sz}, A., \& {Rosotti}, G.~P. 2018, \mnras, 474, L32,
  \dodoi{10.1093/mnrasl/slx182}

\bibitem[{{Keppler} {et~al.}(2018){Keppler}, {Benisty}, {M{\"u}ller},
  {Henning}, {van Boekel}, {Cantalloube}, {Ginski}, {van Holstein}, {Maire},
  {Pohl}, {Samland }, {Avenhaus}, {Baudino}, {Boccaletti}, {de Boer},
  {Bonnefoy}, {Chauvin}, {Desidera}, {Langlois}, {Lazzoni}, {Marleau},
  {Mordasini}, {Pawellek}, {Stolker}, {Vigan}, {Zurlo}, {Birnstiel},
  {Brandner}, {Feldt}, {Flock}, {Girard}, {Gratton}, {Hagelberg}, {Isella},
  {Janson}, {Juhasz}, {Kemmer}, {Kral}, {Lagrange}, {Launhardt}, {Matter},
  {M{\'e}nard}, {Milli}, {Molli{\`e}re}, {Olofsson}, {P{\'e}rez}, {Pinilla},
  {Pinte}, {Quanz}, {Schmidt}, {Udry}, {Wahhaj}, {Williams}, {Buenzli},
  {Cudel}, {Dominik}, {Galicher}, {Kasper}, {Lannier}, {Mesa}, {Mouillet},
  {Peretti}, {Perrot}, {Salter}, {Sissa}, {Wildi}, {Abe}, {Antichi},
  {Augereau}, {Baruffolo}, {Baudoz}, {Bazzon}, {Beuzit}, {Blanchard}, {Brems},
  {Buey}, {De Caprio}, {Carbillet}, {Carle}, {Cascone}, {Cheetham}, {Claudi},
  {Costille}, {Delboulb{\'e}}, {Dohlen}, {Fantinel}, {Feautrier}, {Fusco},
  {Giro}, {Gluck}, {Gry}, {Hubin}, {Hugot}, {Jaquet}, {Le Mignant}, {Llored},
  {Madec}, {Magnard}, {Martinez}, {Maurel}, {Meyer}, {M{\"o}ller-Nilsson},
  {Moulin}, {Mugnier}, {Orign{\'e}}, {Pavlov}, {Perret}, {Petit}, {Pragt},
  {Puget}, {Rabou}, {Ramos}, {Rigal}, {Rochat}, {Roelfsema}, {Rousset}, {Roux},
  {Salasnich}, {Sauvage}, {Sevin}, {Soenke}, {Stadler}, {Suarez}, {Turatto}, \&
  {Weber}}]{Keppler2018}
{Keppler}, M., {Benisty}, M., {M{\"u}ller}, A., {et~al.} 2018, \aap, 617, A44,
  \dodoi{10.1051/0004-6361/201832957}

\bibitem[{{Kraus} {et~al.}(2017){Kraus}, {Kreplin}, {Fukugawa}, {Muto},
  {Sitko}, {Young}, {Bate}, {Grady}, {Harries}, {Monnier}, {Willson}, \&
  {Wisniewski}}]{kraus17}
{Kraus}, S., {Kreplin}, A., {Fukugawa}, M., {et~al.} 2017, \apjl, 848, L11,
  \dodoi{10.3847/2041-8213/aa8edc}

\bibitem[{{Kuhn} {et~al.}(2001){Kuhn}, {Potter}, \& {Parise}}]{Kuhn2001}
{Kuhn}, J.~R., {Potter}, D., \& {Parise}, B. 2001, \apjl, 553, L189,
  \dodoi{10.1086/320686}

\bibitem[{{Lyra} {et~al.}(2016){Lyra}, {Richert}, {Boley}, {Turner}, {Mac Low},
  {Okuzumi}, \& {Flock}}]{Lyra2016}
{Lyra}, W., {Richert}, A. J.~W., {Boley}, A., {et~al.} 2016, \apj, 817, 102,
  \dodoi{10.3847/0004-637X/817/2/102}

\bibitem[{{Marois} {et~al.}(2008){Marois}, {Macintosh}, {Barman}, {Zuckerman},
  {Song}, {Patience}, {Lafreni{\`e}re}, \& {Doyon}}]{Marois2008}
{Marois}, C., {Macintosh}, B., {Barman}, T., {et~al.} 2008, Science, 322, 1348,
  \dodoi{10.1126/science.1166585}

\bibitem[{{Mawet} {et~al.}(2017){Mawet}, {Choquet}, {Absil}, {Huby}, {Bottom},
  {Serabyn}, {Femenia}, {Lebrgeton}, {Matthews}, \& {Gomez
  Gonzalez}}]{Mawet2017}
{Mawet}, D., {Choquet}, {\'E}., {Absil}, O., {et~al.} 2017, \aj, 153, 44,
  \dodoi{10.3847/1538-3881/153/1/44}

\bibitem[{{McDonald} {et~al.}(2017){McDonald}, {Zijlstra}, \&
  {Watson}}]{McDonald2017}
{McDonald}, I., {Zijlstra}, A.~A., \& {Watson}, R.~A. 2017, Monthly Notices of
  the Royal Astronomical Society, 471, 770, \dodoi{10.1093/mnras/stx1433}

\bibitem[{{Milli} {et~al.}(2012){Milli}, {Mouillet}, {Lagrange}, {Boccaletti},
  {Mawet}, {Chauvin}, \& {Bonnefoy}}]{Milli2012}
{Milli}, J., {Mouillet}, D., {Lagrange}, A.~M., {et~al.} 2012, \aap, 545, A111,
  \dodoi{10.1051/0004-6361/201219687}

\bibitem[{{Montesinos} \& {Cuello}(2018)}]{Montesinos2018}
{Montesinos}, M., \& {Cuello}, N. 2018, \mnras, 475, L35,
  \dodoi{10.1093/mnrasl/sly001}

\bibitem[{{Montesinos} {et~al.}(2016){Montesinos}, {Perez}, {Casassus},
  {Marino}, {Cuadra}, \& {Christiaens}}]{Montesinos2016}
{Montesinos}, M., {Perez}, S., {Casassus}, S., {et~al.} 2016, \apjl, 823, L8,
  \dodoi{10.3847/2041-8205/823/1/L8}

\bibitem[{{Muto} {et~al.}(2012){Muto}, {Grady}, {Hashimoto}, {Fukagawa},
  {Hornbeck}, {Sitko}, {Russell}, {Werren}, {Cur{\'e}}, {Currie}, {Ohashi},
  {Okamoto}, {Momose}, {Honda}, {Inutsuka}, {Takeuchi}, {Dong}, {Abe},
  {Brandner}, {Brandt}, {Carson}, {Egner}, {Feldt}, {Fukue}, {Goto}, {Guyon},
  {Hayano}, {Hayashi}, {Hayashi}, {Henning}, {Hodapp}, {Ishii}, {Iye},
  {Janson}, {Kandori}, {Knapp}, {Kudo}, {Kusakabe}, {Kuzuhara}, {Matsuo},
  {Mayama}, {McElwain}, {Miyama}, {Morino}, {Moro-Martin}, {Nishimura}, {Pyo},
  {Serabyn}, {Suto}, {Suzuki}, {Takami}, {Takato}, {Terada}, {Thalmann},
  {Tomono}, {Turner}, {Watanabe}, {Wisniewski}, {Yamada}, {Takami}, {Usuda}, \&
  {Tamura}}]{Muto2012}
{Muto}, T., {Grady}, C.~A., {Hashimoto}, J., {et~al.} 2012, \apjl, 748, L22,
  \dodoi{10.1088/2041-8205/748/2/L22}

\bibitem[{{Natta} {et~al.}(2001){Natta}, {Prusti}, {Neri}, {Wooden}, {Grinin},
  \& {Mannings}}]{Natta2001}
{Natta}, A., {Prusti}, T., {Neri}, R., {et~al.} 2001, \aap, 371, 186,
  \dodoi{10.1051/0004-6361:20010334}

\bibitem[{{Ohta} {et~al.}(2016){Ohta}, {Fukagawa}, {Sitko}, {Muto}, {Kraus},
  {Grady}, {Wisniewski}, {Swearingen}, {Shibai}, {Sumi}, {Hashimoto}, {Kudo},
  {Kusakabe}, {Momose}, {Okamoto}, {Kotani}, {Takami}, {Currie}, {Thalmann},
  {Janson}, {Akiyama}, {Follette}, {Mayama}, {Abe}, {Brandner}, {Brandt},
  {Carson}, {Egner}, {Feldt}, {Goto}, {Guyon}, {Hayano}, {Hayashi}, {Hayashi},
  {Henning}, {Hodapp}, {Ishii}, {Iye}, {Kandori}, {Knapp}, {Kuzuhara}, {Kwon},
  {Matsuo}, {McElwain}, {Miyama}, {Morino}, {Moro-Mart{\'{\i}}n}, {Nishimura},
  {Pyo}, {Serabyn}, {Suenaga}, {Suto}, {Suzuki}, {Takahashi}, {Takami},
  {Takato}, {Terada}, {Tomono}, {Turner}, {Usuda}, {Watanabe}, {Yamada}, \&
  {Tamura}}]{Ohta16}
{Ohta}, Y., {Fukagawa}, M., {Sitko}, M.~L., {et~al.} 2016, \pasj, 68, 53,
  \dodoi{10.1093/pasj/psw051}

\bibitem[{{P{\'e}rez} {et~al.}(2016){P{\'e}rez}, {Carpenter}, {Andrews},
  {Ricci}, {Isella}, {Linz}, {Sargent}, {Wilner}, {Henning}, {Deller},
  {Chandler}, {Dullemond}, {Lazio}, {Menten}, {Corder}, {Storm}, {Testi},
  {Tazzari}, {Kwon}, {Calvet}, {Greaves}, {Harris}, \& {Mundy}}]{Perez2016}
{P{\'e}rez}, L.~M., {Carpenter}, J.~M., {Andrews}, S.~M., {et~al.} 2016,
  Science, 353, 1519, \dodoi{10.1126/science.aaf8296}

\bibitem[{{Pinte} {et~al.}(2018){Pinte}, {Price}, {M{\'e}nard}, {Duch{\^e}ne},
  {Dent}, {Hill}, {de Gregorio-Monsalvo}, {Hales}, \& {Mentiplay}}]{Pinte2018}
{Pinte}, C., {Price}, D.~J., {M{\'e}nard}, F., {et~al.} 2018, \apjl, 860, L13,
  \dodoi{10.3847/2041-8213/aac6dc}

\bibitem[{{Rafikov}(2002)}]{Rafikov2002}
{Rafikov}, R.~R. 2002, \apj, 569, 997, \dodoi{10.1086/339399}

\bibitem[{{Rameau} {et~al.}(2012){Rameau}, {Chauvin}, {Lagrange},
  {Th{\'e}bault}, {Milli}, {Girard}, \& {Bonnefoy}}]{Rameau2012}
{Rameau}, J., {Chauvin}, G., {Lagrange}, A.~M., {et~al.} 2012, \aap, 546, A24,
  \dodoi{10.1051/0004-6361/201219736}

\bibitem[{{Reggiani} {et~al.}(2018){Reggiani}, {Christiaens}, {Absil}, {Mawet},
  {Huby}, {Choquet}, {Gomez Gonzalez}, {Ruane}, {Femenia}, {Serabyn},
  {Matthews}, {Barraza}, {Carlomagno}, {Defr{\`e}re}, {Delacroix}, {Habraken},
  {Jolivet}, {Karlsson}, {Orban de Xivry}, {Piron}, {Surdej}, {Vargas Catalan},
  \& {Wertz}}]{Reggiani2018}
{Reggiani}, M., {Christiaens}, V., {Absil}, O., {et~al.} 2018, \aap, 611, A74,
  \dodoi{10.1051/0004-6361/201732016}

\bibitem[{{Ruane} {et~al.}(2019){Ruane}, {Ngo}, {Mawet}, {Absil}, {Choquet},
  {Cook}, {Gomez Gonzalez}, {Huby}, {Matthews}, \& {Meshkat}}]{Ruane2019}
{Ruane}, G., {Ngo}, H., {Mawet}, D., {et~al.} 2019, \aj, 157, 118,
  \dodoi{10.3847/1538-3881/aafee2}

\bibitem[{{Serabyn} {et~al.}(2017){Serabyn}, {Huby}, {Matthews}, {Mawet},
  {Absil}, {Femenia}, {Wizinowich}, {Karlsson}, {Bottom}, {Campbell},
  {Carlomagno}, {Defr{\`e}re}, {Delacroix}, {Forsberg}, {Gomez Gonzalez},
  {Habraken}, {Jolivet}, {Liewer}, {Lilley}, {Piron}, {Reggiani}, {Surdej},
  {Tran}, {Vargas Catal{\'a}n}, \& {Wertz}}]{Serabyn2017}
{Serabyn}, E., {Huby}, E., {Matthews}, K., {et~al.} 2017, \aj, 153, 43,
  \dodoi{10.3847/1538-3881/153/1/43}

\bibitem[{{Soummer} {et~al.}(2012){Soummer}, {Pueyo}, \&
  {Larkin}}]{Soummer2012}
{Soummer}, R., {Pueyo}, L., \& {Larkin}, J. 2012, The Astrophysical Journal,
  755, L28, \dodoi{10.1088/2041-8205/755/2/L28}

\bibitem[{{Strom} {et~al.}(1989){Strom}, {Strom}, {Edwards}, {Cabrit}, \&
  {Skrutskie}}]{Strom1989}
{Strom}, K.~M., {Strom}, S.~E., {Edwards}, S., {Cabrit}, S., \& {Skrutskie},
  M.~F. 1989, \aj, 97, 1451, \dodoi{10.1086/115085}

\bibitem[{{Tamura}(2009)}]{Tamura2009}
{Tamura}, M. 2009, in American Institute of Physics Conference Series, ed.
  T.~{Usuda}, M.~{Tamura}, \& M.~{Ishii}, Vol. 1158, 11--16,
  \dodoi{10.1063/1.3215811}

\bibitem[{{Tang} {et~al.}(2017){Tang}, {Guilloteau}, {Dutrey}, {Muto}, {Shen},
  {Gu}, {Inutsuka}, {Momose}, {Pietu}, {Fukagawa}, {Chapillon}, {Ho}, {di
  Folco}, {Corder}, {Ohashi}, \& {Hashimoto}}]{Tang2017}
{Tang}, Y.-W., {Guilloteau}, S., {Dutrey}, A., {et~al.} 2017, \apj, 840, 32,
  \dodoi{10.3847/1538-4357/aa6af7}

\bibitem[{{Testi} {et~al.}(2001){Testi}, {Natta}, {Shepherd}, \&
  {Wilner}}]{Testi2001}
{Testi}, L., {Natta}, A., {Shepherd}, D.~S., \& {Wilner}, D.~J. 2001, \apj,
  554, 1087, \dodoi{10.1086/321406}

\bibitem[{{Tsukagoshi} {et~al.}(2019){Tsukagoshi}, {Muto}, {Nomura}, {Kawabe},
  {Kanagawa}, {Okuzumi}, {Ida}, {Walsh}, {Millar}, {Takahashi}, {Hashimoto},
  {Uyama}, \& {Tamura}}]{Tsukagoshi2019}
{Tsukagoshi}, T., {Muto}, T., {Nomura}, H., {et~al.} 2019, \apjl, 878, L8,
  \dodoi{10.3847/2041-8213/ab224c}

\bibitem[{{Ubeira Gabellini} {et~al.}(2019){Ubeira Gabellini}, {Miotello},
  {Facchini}, {Ragusa}, {Lodato}, {Testi}, {Benisty}, {Bruderer}, {T.
  Kurtovic}, \& {Andrews}}]{Ubeira2019}
{Ubeira Gabellini}, M.~G., {Miotello}, A., {Facchini}, S., {et~al.} 2019,
  \mnras, 486, 4638, \dodoi{10.1093/mnras/stz1138}

\bibitem[{{Uyama} {et~al.}(2017){Uyama}, {Hashimoto}, {Kuzuhara}, {Mayama},
  {Akiyama}, {Currie}, {Livingston}, {Kudo}, {Kusakabe}, \& {Abe}}]{Uyama2017}
{Uyama}, T., {Hashimoto}, J., {Kuzuhara}, M., {et~al.} 2017, \aj, 153, 106,
  \dodoi{10.3847/1538-3881/153/3/106}

\bibitem[{{Uyama} {et~al.}(2018){Uyama}, {Hashimoto}, {Muto}, {Akiyama},
  {Dong}, {de Leon}, {Sakon}, {Kudo}, {Kusakabe}, {Kuzuhara}, {Bonnefoy},
  {Abe}, {Brand ner}, {Brandt}, {Carson}, {Currie}, {Egner}, {Feldt}, {Fung},
  {Goto}, {Grady}, {Guyon}, {Hayano}, {Hayashi}, {Hayashi}, {Henning},
  {Hodapp}, {Ishii}, {Iye}, {Janson}, {Kand ori}, {Knapp}, {Kwon}, {Matsuo},
  {Mayama}, {Mcelwain}, {Miyama}, {Morino}, {Moro-Martin}, {Nishimura}, {Pyo},
  {Serabyn}, {Sitko}, {Suenaga}, {Suto}, {Suzuki}, {Takahashi}, {Takami},
  {Takato}, {Terada}, {Thalmann}, {Turner}, {Watanabe}, {Wisniewski}, {Yamada},
  {Yang}, {Takami}, {Usuda}, \& {Tamura}}]{Uyama2018}
{Uyama}, T., {Hashimoto}, J., {Muto}, T., {et~al.} 2018, \aj, 156, 63,
  \dodoi{10.3847/1538-3881/aacbd1}

\bibitem[{{van der Marel} {et~al.}(2013){van der Marel}, {van Dishoeck},
  {Bruderer}, {Birnstiel}, {Pinilla}, {Dullemond}, {van Kempen}, {Schmalzl},
  {Brown}, {Herczeg}, {Mathews}, \& {Geers}}]{vanderMarel2013}
{van der Marel}, N., {van Dishoeck}, E.~F., {Bruderer}, S., {et~al.} 2013,
  Science, 340, 1199, \dodoi{10.1126/science.1236770}

\bibitem[{{Wagner} {et~al.}(2019){Wagner}, {Stone}, {Spalding}, {Apai}, {Dong},
  {Ertel}, {Leisenring}, \& {Webster}}]{Wagner2019}
{Wagner}, K., {Stone}, J.~M., {Spalding}, E., {et~al.} 2019, \apj, 882, 20,
  \dodoi{10.3847/1538-4357/ab32ea}

\bibitem[{{Xuan} {et~al.}(2018){Xuan}, {Mawet}, {Ngo}, {Ruane}, {Bailey},
  {Choquet}, {Absil}, {Alvarez}, {Bryan}, {Cook}, {Femen{\'\i}a Castell{\'a}},
  {Gomez Gonzalez}, {Huby}, {Knutson}, {Matthews}, {Ragland}, {Serabyn}, \&
  {Zawol}}]{Xuan2018}
{Xuan}, W.~J., {Mawet}, D., {Ngo}, H., {et~al.} 2018, \aj, 156, 156,
  \dodoi{10.3847/1538-3881/aadae6}

\bibitem[{{Zhu} {et~al.}(2015){Zhu}, {Dong}, {Stone}, \& {Rafikov}}]{Zhu2015}
{Zhu}, Z., {Dong}, R., {Stone}, J.~M., \& {Rafikov}, R.~R. 2015, \apj, 813, 88,
  \dodoi{10.1088/0004-637X/813/2/88}

\end{thebibliography}

\appendix
\section{Supplementary Keck/NIRC2 Images}
\label{sec: Supplementary Keck/NIRC2 Images}

\begin{figure*}[h]
    \centering
    \includegraphics[scale=0.45]{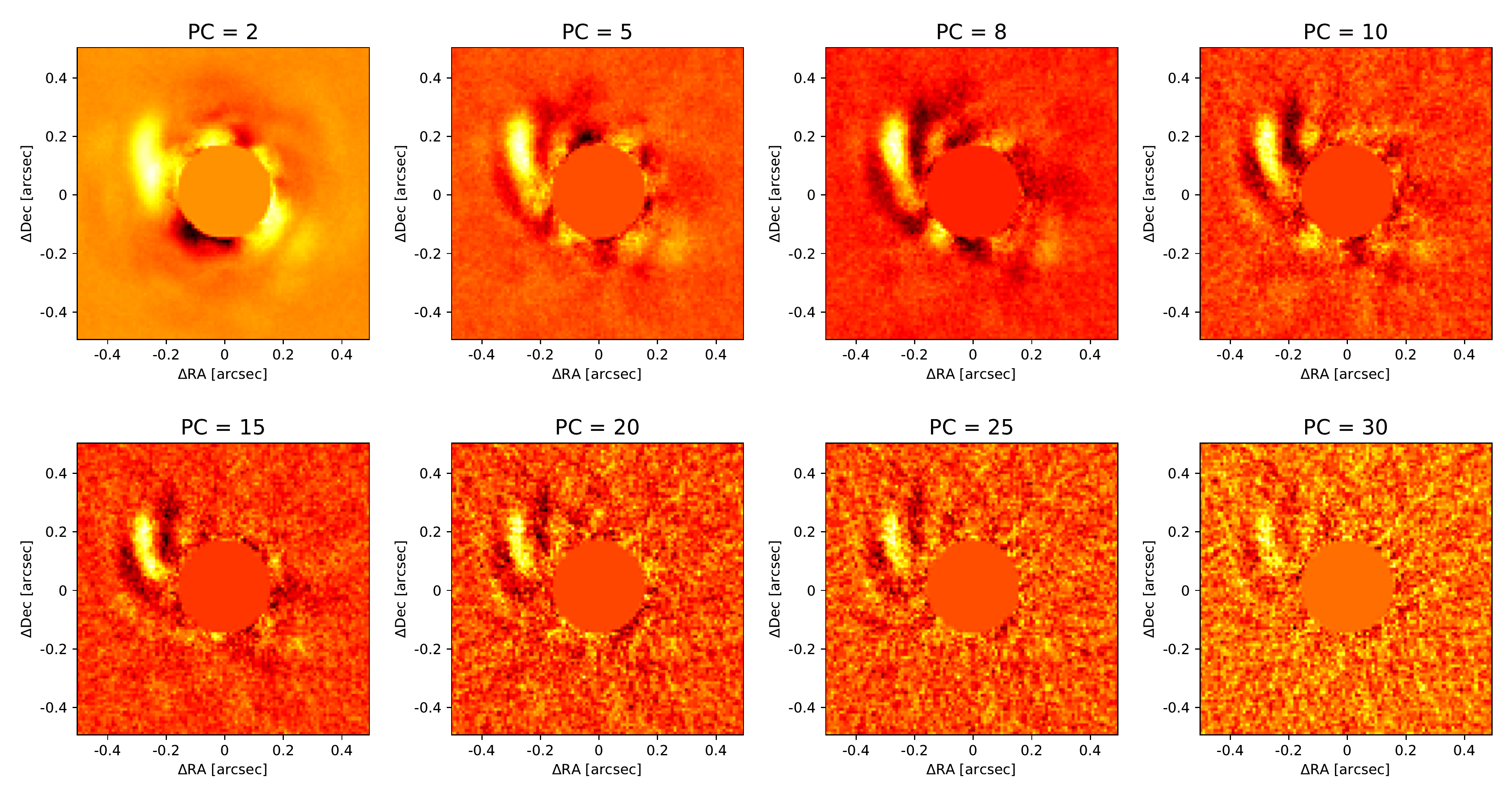}
    \caption{Keck/NIRC2 ADI-reduced images of CQ Tau at different PCs. These images also detected the same spiral structure with SNRs$>$5.}
    \label{fig: NIRC2 PCs}
\end{figure*}

\begin{figure*}
    \begin{tabular}{cc}
    \begin{minipage}{0.5\hsize}
    \centering
    \includegraphics[scale=0.65]{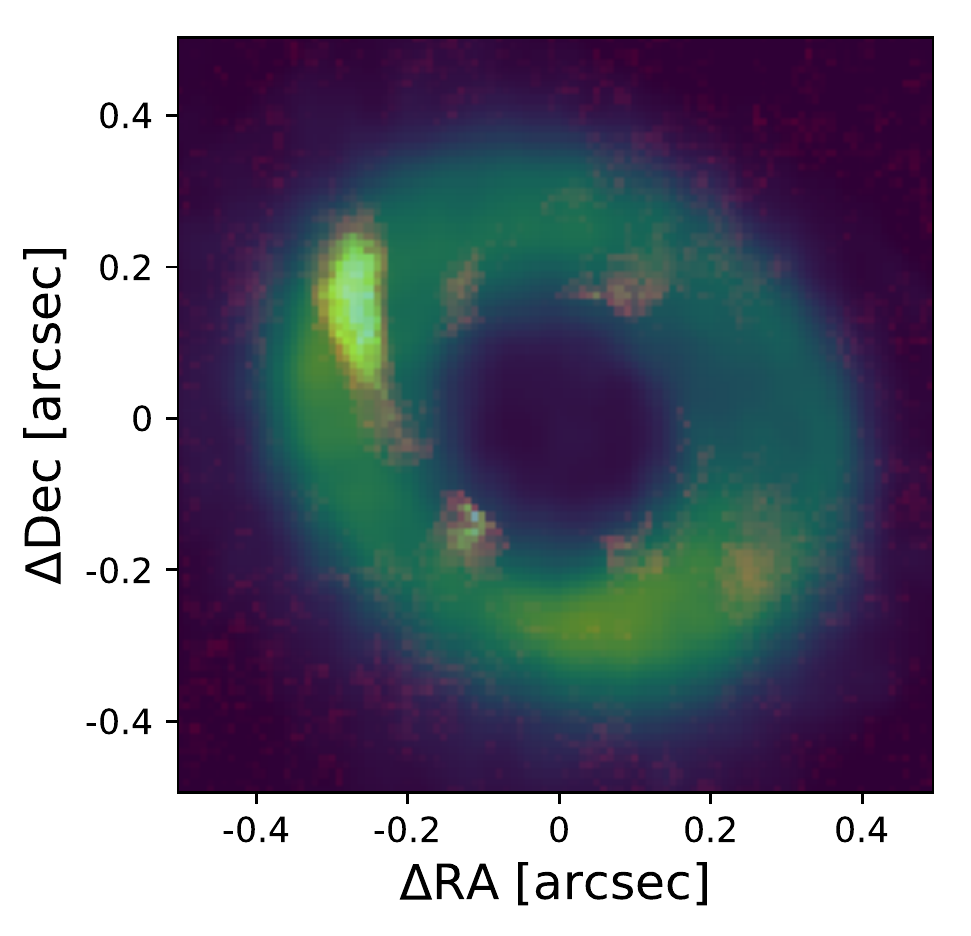}
    \end{minipage}
    \begin{minipage}{0.5\hsize}
    \centering
    \includegraphics[scale=0.65]{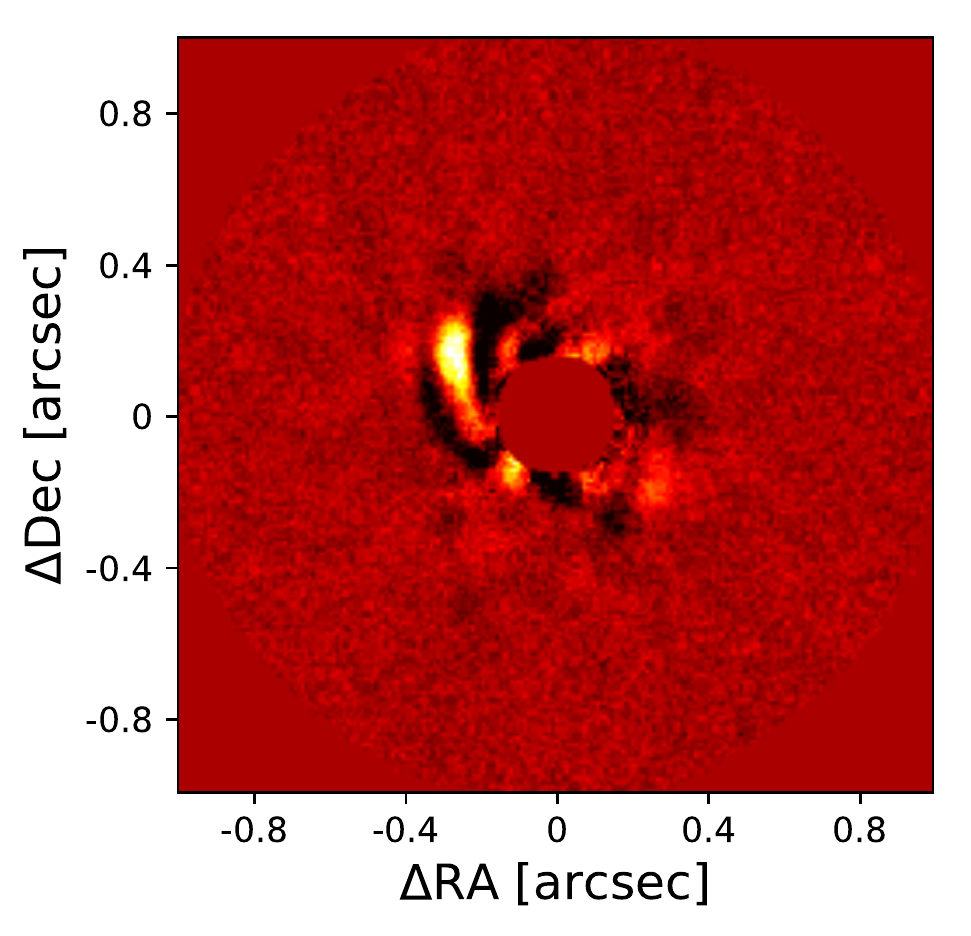}
    \end{minipage}
    \end{tabular}
    \caption{(left) As Figure \ref{fig: NIRC2} superimposed with the ALMA continuum (yellow-green). (right) ADI-reduced NIRC2 image (PC=8) with a larger FoV.}
    \label{fig: NIRC2 supplements}
\end{figure*}

\begin{figure*}
    \centering
    \includegraphics[scale=0.23]{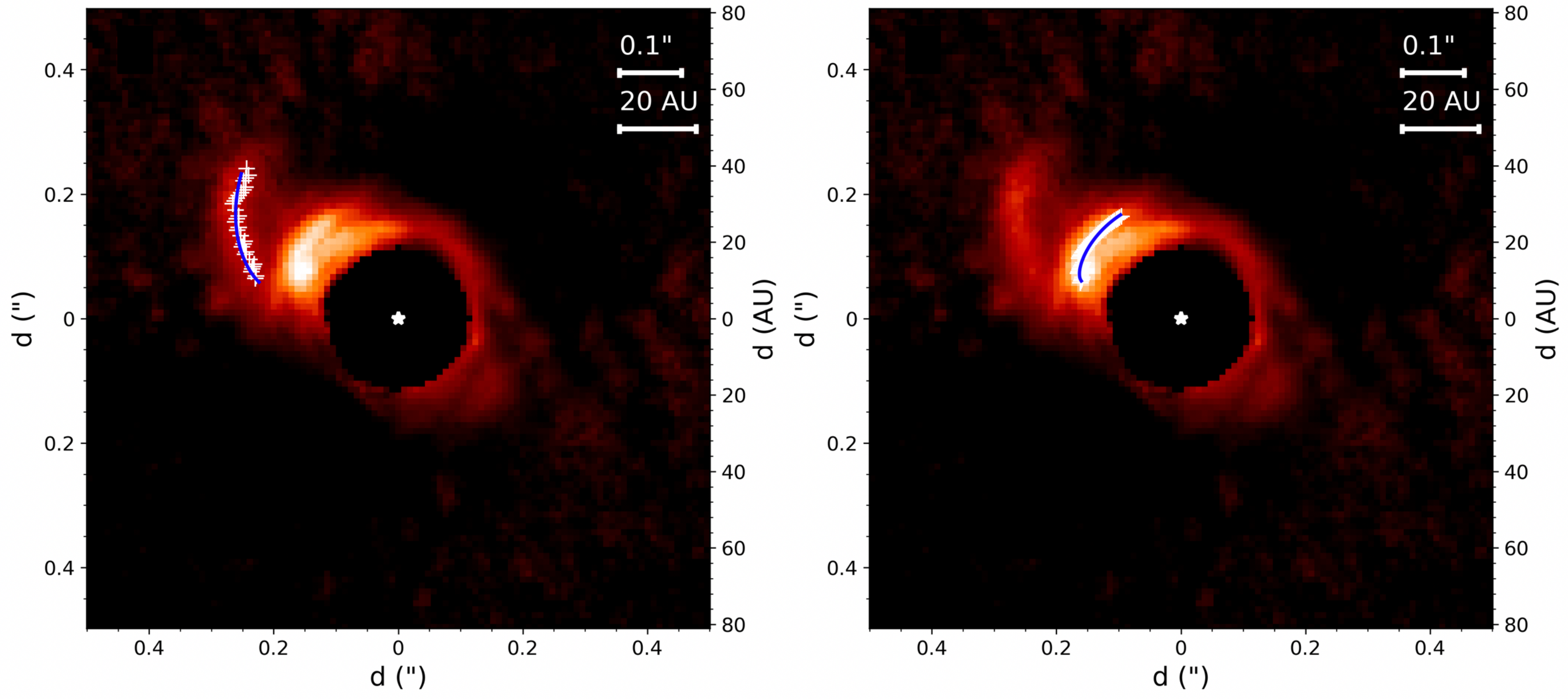}
    \caption{As middle and right images in Figure \ref{fig: HiCIAO} for the fitted result with the best-fit general Archimedean spirals.}
    \label{fig: fit with archimedean}
\end{figure*}

We present supplementary images to clearly show our Keck/NIRC2 result.
Figure \ref{fig: NIRC2 PCs} presents a set of different PCs. Figure \ref{fig: NIRC2 supplements} shows the NIRC2 results superimposed with the ALMA continuum (left) and a full FoV version of the VIP-ADI reduction (right).
Figure \ref{fig: fit with archimedean} shows the best-fit Archimedean spirals ($r=a+b\times\theta^n$) that reproduce well the observed features (left for the outer spiral: $a = 0\farcs221\pm0\farcs004,\ b = 0\farcs203\pm0\farcs010,\ n = 0.744 \pm 0.045$ and right for the inner feature: $a = 0\farcs141 \pm0\farcs006,\ b = 0\farcs056 \pm 0\farcs007,\ n = 0.149 \pm 0.079$).

\end{document}